\documentclass[a4paper,fleqn,usenatbib]{mnras}
\usepackage{amsmath}
\usepackage{bm}
\usepackage{amssymb}
\usepackage{txfonts}
\usepackage{multicol}

\usepackage[T1]{fontenc}
\usepackage{ae,aecompl}
\usepackage{graphicx}	% Including figure files
\graphicspath{{fig/}}	% all figures live in fig/
\usepackage{subfigure}
\usepackage{grffile}
\usepackage{xspace}
\usepackage{epsf}
\usepackage{slashbox}

\AtBeginDocument{}%
\begin{document}
%units:
	\newcommand{\msun}{\,{\rm M}_{\odot}}
\newcommand{\IMBH}{\bullet}
	\newcommand{\rsun}{R_{\odot}}
	\newcommand{\kms}{\, {\rm km\, s}^{-1}}
	\newcommand{\cm}{\, {\rm cm}}
	\newcommand{\gm}{\, {\rm g}}
	\newcommand{\erg}{\, {\rm erg}}
	\newcommand{\kel}{\, {\rm K}}
	\newcommand{\pc}{\, {\rm pc}}
	\newcommand{\kpc}{\, {\rm kpc}}
	\newcommand{\mpc}{\, {\rm Mpc}}
	\newcommand{\seg}{\, {\rm s}}
	\newcommand{\kev}{\, {\rm keV}}
	\newcommand{\hz}{\, {\rm Hz}}
	\newcommand{\etal}{et al.\ }
	\newcommand{\yr}{\, {\rm yr}}
	\newcommand{\gyr}{\, {\rm Gyr}}
	\newcommand{\eq}{eq.\ }
	\newcommand{\amunit}{\msun {\rm AU^2/yr}}
	\def\arcsec{''\hskip-3pt .}
    % more useful commands from Kocsis & Tremaine (2015)
    \def\half{{\textstyle{\frac{1}{2}}}}
    \def\gothg{\,\mathfrak G}
    \def\ffrac#1#2{{\textstyle\frac{#1}{#2}}}
    \newcommand{\kmsf}{\frac{\rm km}{\rm s}}
    \newcommand{\Myr}{\,{\rm Myr}}
    \newcommand{\Gyr}{\,{\rm Gyr}}
    \newcommand{\D}{\mathrm{d}}
    \newcommand{\Msun}{\,\mathrm{M}_{\odot}}
    \newcommand{\E}{\mathrm{VRR}}
    \newcommand{\tot}{\mathrm{tot}}
    \newcommand{\J}{\mathcal{J}}
    \newcommand{\Ln}{\bm{\hat{\L}}}
    \renewcommand{\L}{\bm{L}}
    \newcommand{\T}{\bm{T}}
    \renewcommand{\O}{\mathbf{O}}
    \newcommand{\G}{\gothg}
    \newcommand{\In}{\rm in}
    \newcommand{\Out}{\rm out}
    \newcommand{\rms}{\rm RMS}
    \renewcommand{\r}{\bm{r}}
    \newcommand{\sat}{\mathrm{sat}}
    \newcommand{\vrr}{\mathrm{vrr}}	
	% symbols used
	\def\gapprox{\;\rlap{\lower 3.0pt                       % approximately smaller
			\hbox{$\sim$}}\raise 2.5pt\hbox{$>$}\;}
	\def\lapprox{\;\rlap{\lower 3.1pt                       % approximately smaller
			\hbox{$\sim$}}\raise 2.7pt\hbox{$<$}\;}
	
	% special case for 4 plots in panels across the page
	\newcommand{\figsizeFour}{9.0cm}
	
	% widths for figures for single and double columns
	\newcommand{\figwidthSingle}{14.0cm}
	\newcommand{\figwidthDouble}{7.50cm}
	
	% ALTER these if/when change from single to double columns
	\newcommand{\figbig}{\figwidthSingle}
	\newcommand{\figsmall}{\figwidthDouble}

	\newcommand{\figappend}{\figwidthSingle}
	
	% own commands for ref equations
	\newcommand{\reqOne}[1]{equation~(\ref{#1})}
	\newcommand{\reqTwo}[2]{equations~(\ref{#1}) and~(\ref{#2})}
	\newcommand{\reqNP}[1]{equation~\ref{#1}}
	\newcommand{\reqTwoNP}[2]{equations~\ref{#1} and~\ref{#2}}
	\newcommand{\reqTo}[2]{equation~(\ref{#1})-(\ref{#2})}
	% equation based commands
	\newcommand{\rn}[1]{(\ref{#1})}
	\newcommand{\ern}[1]{equation~(\ref{#1})}
	\newcommand{\be}{\begin{equation}}
	\newcommand{\ee}{\end{equation}}
	\newcommand{\ff}[2]{{\textstyle \frac{#1}{#2}}}
	\newcommand{\ben}{\begin{enumerate}}
		\newcommand{\een}{\end{enumerate}}
% text color comments:
	\newcommand{\tred[1]}{\textcolor{red}{#1}}
	\newcommand{\tgreen[1]}{\textcolor{green}{#1}}
\def\BK#1{{\color{red}\bf BK: #1}}
\defcitealias{Panamarev2026}{Paper~I}
\newcommand{\PaperI}{\citetalias{Panamarev2026}\xspace}

\newcommand{\tp}[1]{\textcolor{blue}{$\langle$\textbf{TP: #1}$\rangle$}}

\title[Stellar discs and IMBHs in galactic nuclei]
  {Stellar discs and intermediate-mass black holes in galactic nuclei --\\II. Fragmenting the disc and forming the S-stars in a live nuclear star cluster}

\author[Panamarev \& Kocsis]
  {Taras Panamarev$^{1,2}$\thanks{Corresponding author email: panamarevt@gmail.com}, and Bence Kocsis$^{1,3}$ \\
   \\
      $^1$ Rudolf Peierls Centre for Theoretical Physics, Parks Road, OX1 3PU, Oxford, UK\\
      $^2$ Fesenkov Astrophysical Institute, Observatory 23, 050020 Almaty, Kazakhstan\\
      $^3$ Institute of Physics, Eötvös University, Pázmány P. s. 1/A, Budapest 1117, Hungary
      }

\maketitle

\begin{abstract}
We present direct $N$-body simulations of a stellar disc around the supermassive black hole Sgr~A*, perturbed by an intermediate-mass black hole (IMBH) and embedded in a live nuclear star cluster. A retrograde IMBH with $m_\IMBH=2000\msun$, relative inclination $\theta_\IMBH=150^\circ$, and initial semi-major axis $a_\IMBH=0.15$~pc  breaks the disc into three disc-like components and an angular momentum filament. The stochastic torques of the cluster, together with the IMBH's inward migration, increase the angular separation between these substructures. By ${\sim}5$~Myr, their separations become comparable to those among the observed Galactic-centre young star structures. The innermost ($a\le0.04$~pc) disc stars evolve into an S-star-like population. Relaxation driven by the cluster disperses their orbital planes, excites their eccentricities and depletes the `zone of avoidance' within ${\sim}10$~Myr even without an IMBH. The IMBH further accelerates their orbital-plane reorientation. However, the most eccentric S-stars may require an additional formation channel, such as the Hills mechanism. The distinctive signatures of the IMBH are the well-separated disc components and the eccentricity excitation confined to the radial range swept by its orbit. Dynamical friction from the cluster reduces the orbital size of a $2000\msun$ IMBH initially at $0.15$~pc by more than a factor of two within $10$~Myr. Prograde IMBHs tend to align with the stellar disc, whereas a $2000\msun$ retrograde IMBH can anti-align with the radially overlapping disc stars. We also quantify how the IMBH affects the supply of stars to the loss cone and the rates of stellar tidal disruptions and binary separations.
\end{abstract}


%###########################################################################
\section{Introduction}\label{sec:INTRO}
%###########################################################################

The central ${\sim}0.5$~pc of the Milky Way harbours about two hundred young, massive stars whose orbits record the recent dynamical history of the Galactic centre \citep{PaumardEtAl2006, Gillessen2017, vonFellenberg2022}. They were originally associated with a single clockwise stellar disc extending from ${\sim}0.04$ to $0.5$~pc around Sgr~A*, identified from the distribution of the stars' orbital angular-momentum directions \citep{LevinBeloborodov2003, PaumardEtAl2006, Yelda2014}. The disc is warped \citep{Bartko2009, vonFellenberg2022}, and its stars move on moderately eccentric orbits ($\langle e \rangle \simeq 0.3$--$0.4$; \citealt{Yelda2014, Siyao2023, vonFellenberg2022}). Recent analyses resolve the young population into several angular-momentum overdensities, including retrograde structures and filaments. Most young stars belong to one of these substructures \citep{Ali2020, vonFellenberg2022, Siyao2023}.

Interior to ${\sim}0.04$~pc resides the S-star cluster, which consists mostly of B-type stars on nearly isotropically oriented orbits with a thermal eccentricity distribution \citep{Gillessen2017}. These stars appear to avoid a well-defined region of the pericentre--eccentricity plane, the `zone of avoidance' \citep{Burkert2024}. 

The two populations appear coeval. The disc formed a few Myr ago ($6\pm2$~Myr, \citealt{PaumardEtAl2006}; or $2.5$--$5.8$~Myr, \citealt{Lu2013}), presumably in situ through the fragmentation of a gaseous accretion disc \citep{LevinBeloborodov2003, Levin2007}, and the S-star ages are consistent with the same star-formation episode \citep{Habibi2017}. Numerical models have also demonstrated the formation of a central stellar disc through star--gas interactions \citep{Panamarev2018}. A disc born this way should be thin, nearly circular, and dynamically cold. The kinematic complexity seen today -- the warp, the substructure, and the isotropic, eccentric innermost cluster -- must therefore have developed within the lifetime of the young population. Whether internal relaxation alone can account for this complexity remains unclear.

Part of the answer lies in the resonant dynamics of the host nuclear star cluster. In the near-Keplerian potential of the supermassive black hole (SMBH), stellar orbits are slowly precessing ellipses that exert coherent mutual torques, so the orbital angular momenta relax much faster than the orbital energies -- the process of scalar and vector resonant relaxation \citep{Rauch1996}. Vector resonant relaxation (VRR) reorients orbital planes on Myr timescales \citep{Hopman2006, Kocsis2015}, bends an embedded disc into a warp comparable to the observed one \citep{Kocsis2011}, and drives the heaviest bodies into flattened, disc-like configurations \citep{Szolgyen2018,Magnan2021,Mathe2022}. Direct $N$-body models of the Milky Way's nuclear star cluster follow these collisional processes self-consistently and have also been used to quantify the rate at which relaxation feeds stars to tidal disruption by the SMBH \citep{Panamarev2019}. In \citet{Panamarev2022}, we evolved a stellar disc inside a live nuclear cluster and found that VRR governs the evolution of its orbital planes. For stars initially in a single disc, those models suggested that reproducing the observed S-star angular-momentum scatter within $10$~Myr would require a $10^{3}\msun$ cusp of stellar-mass black holes or an intermediate-mass black hole (IMBH). Similar evolution occurs in globular clusters, where stellar discs dissolve within a local relaxation time while retaining a kinematic memory of their origin \citep{MastrobuonoBattisti2013, Meiron2019, Berczik2025}.

This motivates the possibility that an additional massive perturber plays an important role in the Galactic centre. In particular, an IMBH can exert strong coherent torques on the young stellar population and drive changes in orbital orientations and eccentricities beyond those produced by the stellar background alone \citep{Perets2007}.
The classical route to the S-stars is the tidal disruption of stellar binaries \citep{Hills1988}, which leaves stars on tight, highly eccentric orbits; their subsequent relaxation can reproduce the observed semi-major-axis and eccentricity distributions, including the zone of avoidance \citep{Perets2009a, Generozov2020, Generozov2025}. The hypervelocity star S5-HVS1, ejected from the Galactic centre ${\sim}5$~Myr ago, demonstrates that this channel has operated recently \citep{Koposov2020}. Alternatively, the S-stars may have migrated inwards through their natal disc \citep{Levin2007}, although the initially nearly circular orbits of migrated stars may fail to reach the observed eccentricities within the stellar ages \citep{Perets2009a, Panamarev2022}. A single massive companion offers a more direct route. An inspiralling IMBH can drag young stars inwards and strip them, leaving a thin disc \citep{Levin2005}. A bound intermediate-mass companion secularly excites the eccentricities of disc stars \citep{Yu+2007,Zheng2020}. An inclined IMBH aligns with the disc through resonant dynamical friction, warping and thickening it during alignment \citep{Szolgyen2021, Ginat2023}. An IMBH inside the S-cluster efficiently randomises the S-star orbits and increases their eccentricities \citep{Merritt+2009,GualandrisMerritt2009}. Single, more energetic events have also been invoked. The gravitational-wave recoil of a recent black hole merger with Sgr~A* would turn the surrounding cluster into an eccentric disc that feeds stars into the innermost region on eccentric, highly inclined orbits \citep{Akiba2025}, and drag from a past accretion disc has been proposed to shape the S-star kinematics as well \citep{Fan2024}. Most recently, \citet{Zheng2026} proposed a unified scenario in which the S-stars, the clockwise disc, and the off-disc stars all descend from a single natal disc shaped by a more massive (${\sim}10^{4}\msun$) intermediate-mass companion orbiting outside the disc, aided by a secular resonance that sweeps through the system as the parent gaseous disc is depleted.

Is such a companion plausible in the Galactic centre? IMBHs ($10^{2}$--$10^{5}\msun$; see \citealt{Greene2020} for a review) may form through the growth of Population~III remnant \citep{Madau_Rees2001}, fragmentation and accretion in an active galactic nucleus accretion disk \citep{Kocsis+2011,McKernan2012,McKernan+2014}, and runaway collisions in dense young clusters or galactic nuclei \citep{Portegies_Zwart+2006, OLeary+2006, Rizzuto2021, Giersz+2015, Rose2022}. If formed externally, infalling star clusters can deliver them to the innermost parsecs and contribute to building the nuclear star cluster \citep{Tremaine1976, Capuzzo-Dolcetta2008, Ishchenko2023, Kuvatova2026}. In the Galactic centre, the IRS13E complex has been argued to host a ${\sim}10^{4}\msun$ black hole \citep{Tsuboi2017, Tsuboi2020}, although this interpretation is contested. $N$-body models show that an IRS13-sized association cannot remain gravitationally bound at its physical distance of $0.4$~pc even with an IMBH of thousands of solar masses, and that the velocity dispersion of a tidally disrupted cluster is accounted for by the tidal field of the SMBH alone \citep{Pavlik2024}; JWST/MIRI monitoring detects no variable mid-infrared counterpart, ruling out an IMBH above ${\sim}10^{3}\msun$ there if it accretes at more than $10^{-6}$ of the Eddington rate \citep{Roychowdhury2025}. The kinematics of a compact molecular cloud farther out suggest a ${\sim}10^{5}\msun$ point mass \citep{Kaneko2023}. The dynamical limits are tightest close to Sgr~A*. The absence of a reflex `wobble' in its proper motion excludes IMBHs more massive than ${\sim}3\times10^{4}\msun$ between $0.003$ and $0.1$~pc \citep{Reid2020}, tightening to ${\sim}3\times10^{3}\msun$ at $0.01$~pc \citep{Oyama2024}. Interferometric astrometry of multiple stellar orbits limits any extended mass inside the apocentre of S2 (${\simeq}9$~mpc) to ${\lesssim}3000\msun$ \citep{Gravity2022}. Analyses of the S2 orbit exclude $10^{3}$--$10^{5}\msun$ companions over a wide range of separations, with the tightest bounds reaching a few hundred solar masses \citep{Gualandris2010, Gravity2023, Naoz2020, Will2023}. The observed hypervelocity stars exclude companions above ${\sim}10^{3}\msun$ within $1$~mpc of Sgr~A* \citep{Evans2023}. Outside the S-star cluster these bounds weaken rapidly, so an IMBH of a few hundred to a few thousand solar masses at the radii occupied by the stellar disc remains permitted -- precisely the regime explored in this work.

This paper investigates how an IMBH shapes the young stellar structures around Sgr~A*. Theoretical predictions for this interaction can be formulated in the space of angular-momentum vector directions within the VRR framework. In \citet{Panamarev2025} we showed that nearly aligned orbits interact most strongly and derived the critical mutual inclination at which a massive perturber disrupts a pair of such orbits, providing an angular-momentum analogue of the Hill radius. We applied this criterion to an initially flat stellar disc to identify configurations capable of producing the observed retrograde structures in the Galactic centre. That criterion assumes the perturber dominates the angular momentum of the system, an assumption that breaks down once the disc is as massive as the IMBH. In \citet{Panamarev2026} (hereafter \PaperI) we relaxed this assumption and derived analytical conditions for the disruption of the stellar disc by a massive perturber. We also numerically surveyed the IMBH--disc interaction embedded in a static background potential using direct $N$-body simulations. While computationally efficient, that approximation suppressed both dynamical friction on the IMBH and stochastic torques from the surrounding stellar cluster. Here, we remove this limitation by modelling the nuclear star cluster (NSC) self-consistently with $10^5$ live particles. Comparison with the model without an IMBH in \citet{Panamarev2022} distinguishes the features produced by the IMBH from those generated by the cluster's internal relaxation. We examine whether an IMBH can explain the orientations and eccentricities of the orbits of the young stars in the Galactic centre, including the S-stars, the clockwise disc, the counterclockwise (retrograde) structure and the other observed angular-momentum substructures.

The paper is organised as follows. Section~\ref{sec:simulations} describes the numerical setup and the initial conditions. Section~\ref{sec:imbh-evol} follows the evolution of the inclination, semi-major axis and eccentricity of the IMBH. Section~\ref{sec:disc-imbh} quantifies how the IMBH thickens and warps the stellar disc, breaks it into substructures in angular-momentum space and excites the eccentricities of its stars. Section~\ref{sec:inner} compares the innermost simulated stars with the observed S-star cluster, including the zone of avoidance, and Section~\ref{sec:tde} presents rates of tidal disruptions, binary separations and Hills captures. Section~\ref{sec:SUM} summarises and discusses our findings.

%###########################################################################
\section{Simulations and initial conditions}
\label{sec:simulations}
%###########################################################################

Our model comprises a central SMBH, an NSC and a stellar disc. We add an IMBH to this system to study its dynamical interaction with the other components. The NSC is represented by a live, self-consistent $N$-body system.\footnote{In \PaperI it was modelled by a smooth isotropic Plummer potential.} The stochastic gravitational fluctuations of the cluster and the associated relaxation processes are followed directly. Below we summarise the relevant aspects of the setup and refer to \PaperI and \citet{Panamarev2022} for a complete description.

\begin{table}
\caption{List of the simulations analysed in this work.}
\label{tab:runs}
\centering
\begin{tabular}{ccccc}
\hline
$\theta_\IMBH$ [$^\circ$] & $a_\IMBH$ [pc] & $e_\IMBH$ & $m_\IMBH$ [${\rm M}_{\odot}$] & $m_\IMBH/M_\mathrm{d}$ \\
\hline
45  & 0.05 & 0.33 & 500  & 0.17 \\
45  & 0.15 & 0.33 & 2000 & 0.67 \\
90  & 0.15 & 0.33 & 2000 & 0.67 \\
135 & 0.05 & 0.33 & 500  & 0.17 \\
135 & 0.15 & 0.70 & 2000 & 0.67 \\
150 & 0.15 & 0.33 & 2000 & 0.67 \\
179 & 0.05 & 0.33 & 2000 & 0.67 \\
\hline
\end{tabular}
\par\medskip
\begin{flushleft}\textbf{Notes.} $\theta_\IMBH$ is the initial inclination of the IMBH orbit relative to the plane of the stellar disc, in degrees; $a_\IMBH$, the initial semi-major axis in parsecs; $e_\IMBH$, the initial eccentricity; $m_\IMBH$, the IMBH mass in solar masses; and $m_\IMBH/M_\mathrm{d}$, the ratio of the IMBH mass to the total disc mass ($M_\mathrm{d}\simeq3000\msun$). A model without an IMBH \citep{Panamarev2022} is used as a baseline for comparison.\end{flushleft}
\end{table}

We use a modified version of the direct $N$-body code phi-GRAPE \citep{HarfstEtAl2007}, which employs a fourth-order Hermite integration scheme and runs on modern GPUs \citep{Gaburov2009}. Our version includes the gravitational interaction between the stars and a massive central object, implemented as an external point-mass potential, as well as the accretion of stars onto that object \citep{JustEtAl2012, KenEtAl2016, Panamarev2019}. The equation of motion of star $i$ is
\begin{equation}\label{eq:motion}
\ddot{\vec r}_{i}=-\sum_{i\neq j}\frac{Gm_{j}\,\vec r_{ij}}
{\left(r_{ij}^{2}+\epsilon_\mathrm{ss}^{2}\right)^{3/2}}
-\frac{G M\,\vec r_{i}}{r_{i}^{3}}\,,
\end{equation}
where $\vec r_{ij}=\vec r_{i}-\vec r_{j}$, with $\vec r_{i}$ and $\vec r_{j}$ the
positions of stars $i$ and $j$, $M$ is the mass of the SMBH, and $\epsilon_\mathrm{ss}=5\times10^{-5}$~pc is a softening length that prevents numerical singularities during close encounters. A star is accreted when its distance from the central object falls below a fixed capture radius, $r_\mathrm{acc}=6.9\times10^{-5}$~pc ($14.2$~AU), and its mass is added to the central object. We refer to these accretion events as tidal disruption events and discuss the physical interpretation of the capture radius in Section~\ref{sec:tde}.
We adopt units in which $G=M=R_\mathrm{out}=1$, where $R_\mathrm{out}$ is the outer edge of the system defined by the semi-major axis of the outermost star. We convert to physical units by setting
$R_\mathrm{out}=0.5$~pc and $M=4\times10^{6}\msun$, appropriate for Sgr~A*.

The NSC comprises $10^{5}$ particles on Keplerian orbits about the SMBH, distributed according to a relaxed, Bahcall--Wolf-like density cusp, $\rho\propto r^{-7/4}$ \citep{Bahcall1976}, corresponding to $\D N/\D a\propto a^{1/4}$ for $10^{-4}\lesssim a\leq0.5$~pc, with a Kroupa top-heavy stellar mass function (Eq.~15 of \citealt{Panamarev2022}) with masses corresponding to an evolved population of stars at 1 Gyr. The cluster has a total mass $M_{*}\approx2\times10^{5}\msun$ and has a thermal eccentricity distribution. 

The stellar disc is the \textsc{stardisc} model of \citet{Panamarev2022}, with $\approx10^{3}$ stars of total mass $M_\mathrm{d}\simeq3000\msun$ on initially nearly circular orbits, as expected for a disc formed through
star--gas interactions in a previously active nucleus \citep{Panamarev2018} or through the fragmentation of a gaseous accretion disc \citep{LevinBeloborodov2003, Levin2007}. The disc follows a three-dimensional density profile $\rho\propto r^{-2.4}$, corresponding to $\D N/\D a\propto a^{-0.4}$ for $0.004\lesssim a\leq0.5$~pc, 
and a top-heavy modification of the \citet{Kroupa2001} initial mass function ($\D N/\D m\propto m^{-1.3}$ above $1\msun$; Eq.~16 of \citealt{Panamarev2022}), with masses evolved to 1~Gyr as for the NSC. The simulations start once the parent gas is fully depleted.

The IMBH is a single, more massive particle subject to the same gravitational interactions as the other particles. We vary its initial inclination relative to the disc plane, its eccentricity, semi-major axis and mass (Table~\ref{tab:runs}). The IMBH could migrate to the Galactic centre from a core of a disrupted globular cluster \citep[see e.g.][]{Tremaine1976, Ishchenko2023, Kuvatova2026} 
or even form in-situ \citep{Rose2022}. We assume that the IMBH was already present in the Galactic centre when the stellar disc formed. 
Therefore, the set of initial conditions in Table~\ref{tab:runs} is intended to span the range of orbits on which the formation channels above could plausibly have delivered an IMBH to the central parsec.

%#############
\section{Evolution of the IMBH}
\label{sec:imbh-evol}
%#############

Compared with the smooth background potential of \PaperI, the live sphere changes the dynamics of the IMBH in two ways. Its stochastic torques perturb the direction of the orbital angular momentum of the IMBH, and dynamical
friction against the cluster stars removes orbital energy and shrinks the semi-major axis of the IMBH over time. In this section we quantify both effects -- first the evolution of the orbital inclination (Section~\ref{subsec:imbh-inc}), then that of the semi-major axis and eccentricity (Section~\ref{subsec:imbh-ae}).

Following \citet{Szolgyen2021} and \PaperI, we categorise the stars by their orbits relative to the IMBH. Therefore, for the disc stars, we define:
\begin{equation}
\label{eq:orb-cond}
\begin{aligned}
&\text{Inner:} \quad && r_{\rm a,*} < r_{\rm p,\IMBH}, \\
&\text{Overlapping:} \quad && r_{\rm a,*} \geq r_{\rm p,\IMBH} \text{~and~} r_{\rm p,*} \leq r_{\rm a,\IMBH}, \\
&\text{Outer:} \quad && r_{\rm p,*} > r_{\rm a,\IMBH},
\end{aligned}
\end{equation}
where $r_{\rm a,*}$ and $r_{\rm p,*}$ are apocentre and pericentre of the disc stars, $r_{\rm a,\IMBH}$ and $r_{\rm p,\IMBH}$ are those of the IMBH.
These conditions are computed at each moment in time.

\subsection{Inclination}
\label{subsec:imbh-inc}

\begin{figure*}
	\begin{centering}
		\includegraphics[width=0.85\linewidth]{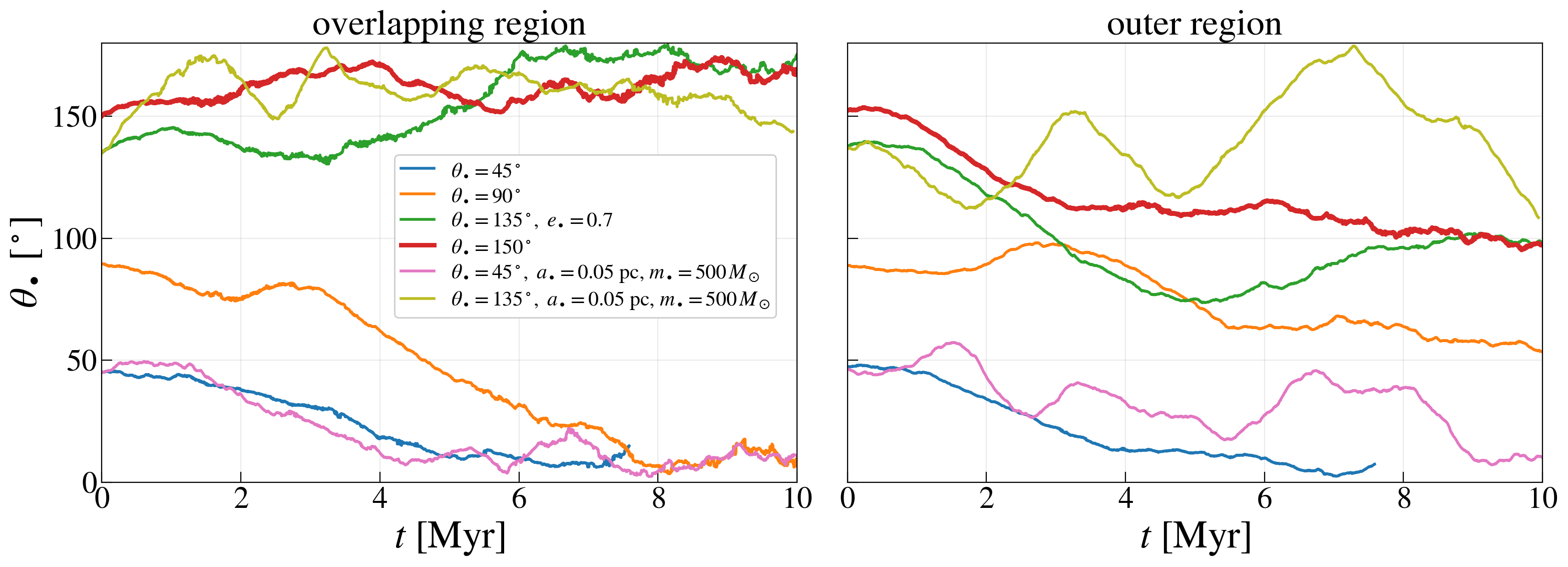} \\
		\par\end{centering}
	\caption{Inclination of the IMBH's orbit as a function of time for the four
	massive ($m_\IMBH=2000\msun$, $a_\IMBH=0.15$~pc) models together with the two low-mass ($m_\IMBH=500\msun$, $a_\IMBH=0.05$~pc) models of
	Table~\ref{tab:runs}. The left and right panels show the angle $\theta_\IMBH$ between the IMBH angular momentum and that of the overlapping and outer disc groups, respectively (Eq.~\eqref{eq:orb-cond}).
	Each curve corresponds to one model, coloured consistently throughout the paper.}
	\label{fig:i_100k}
\end{figure*}

\begin{figure*}
	\begin{centering}
		\includegraphics[width=0.85\linewidth]{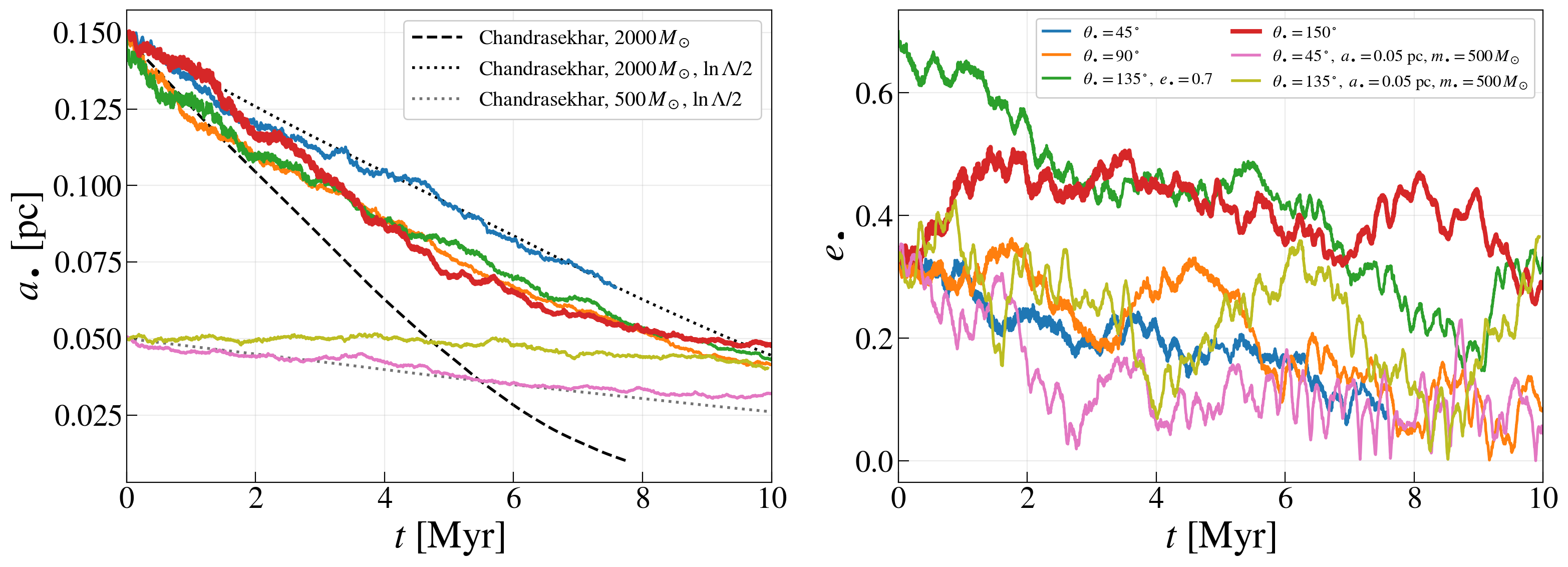} \\
		\par\end{centering}
	\caption{Semi-major axis $a_\IMBH$ (left) and eccentricity $e_\IMBH$
	(right) of the IMBH as a function of time, for the same models and with the
	same colours as in Fig.~\ref{fig:i_100k}. The black dashed and dotted
	curves in the left panel show the Chandrasekhar dynamical-friction
	prediction of Eq.~\eqref{eq:df} for the massive ($m_\IMBH=2000\msun$,
	$a_\IMBH(0)=0.15$~pc) models, with the standard
	$\ln\Lambda=\ln(M/m_\IMBH)$ and with the fitted
	$\ln\Lambda=\frac{1}{2}\ln(M/m_\IMBH)$, respectively; the grey dotted
	curve shows the $\ln\Lambda=\frac{1}{2}\ln(M/m_\IMBH)$ prediction for the
	low-mass ($m_\IMBH=500\msun$, $a_\IMBH(0)=0.05$~pc) models.}
	\label{fig:a_e_100k}
\end{figure*}

Figure~\ref{fig:i_100k} shows the inclination of the IMBH's orbit with respect to the disc as a function of time. For the prograde models the behaviour is similar to that found with the fixed spherical potential of \PaperI. Both the massive $\theta_\IMBH=45^\circ$ model and its low-mass counterpart align with the radially overlapping stars within $\sim$5--8~Myr (left panel of Fig.~\ref{fig:i_100k}), as expected from resonant dynamical friction \citep{Szolgyen2021, Ginat2023}. With respect to the outer disc (right panel) the massive prograde model aligns as well; the low-mass one also tends
towards alignment, but oscillates about the outer disc's angular-momentum direction along the way.

The retrograde IMBHs with $m_\IMBH=2000\msun=0.67M_{\rm d}$ instead tend to anti-align with the radially overlapping stars, similarly to the behaviour seen in simulations using the smooth spherical Plummer potential (\PaperI). With respect to the outer disc the IMBHs tend towards an orthogonal configuration by the end of their runs. 

In \PaperI the orthogonal configuration was stationary, because the mutual torques between the IMBH and the disc vanish at $\theta_\IMBH=90^\circ$. In the live sphere this equilibrium is not maintained, and the initially orthogonal model (orange curves in Fig.~\ref{fig:i_100k}) aligns with the radially overlapping stars within ${\sim}8$~Myr while remaining inclined by ${\sim}50^\circ$ with respect to the outer disc. We attribute the breaking of this marginal equilibrium primarily to the stochastic torques exerted by the live cluster, which are absent from the smooth background and perturb the IMBH away from the $\theta_\IMBH=90^\circ$ configuration. In addition, the geometry itself evolves, since dynamical friction carries the IMBH inward and thereby changes the population of radially overlapping stars against which the relative inclination is defined.

\subsection{Semi-major axis and eccentricity}
\label{subsec:imbh-ae}

We next analyse the semi-major axes (left panel of Fig.~\ref{fig:a_e_100k}).
For the $m_\IMBH=2000\msun$ IMBHs initially at $a_\IMBH=0.15$~pc, the semi-major axis shrinks by more than a factor of two (up to about $70$\%) by $10$~Myr as the IMBH sinks towards the SMBH. The decay proceeds at broadly similar rates in all four models irrespective of their orbital inclination with respect to the disc, as expected when the dynamical-friction drag is dominated by the isotropic, non-rotating cluster. The $500\msun$ low-mass IMBHs migrate far less, the retrograde one changing its semi-major axis by less than $20$\% over the same interval.

The dashed and dotted curves in the left panel of Fig.~\ref{fig:a_e_100k} compare the measured decay with the prediction of Chandrasekhar's formula for a body on a circular orbit,
\begin{equation}\label{eq:df}
\frac{\D a}{\D t}=-\,\frac{8\pi G\,m_\IMBH\,\rho(a)\,\ln\Lambda\,f(X)\,a^{2}}{M\,v_\mathrm{c}(a)}\,,
\qquad
f(X)=\mathrm{erf}(X)-\frac{2X\,\mathrm{e}^{-X^{2}}}{\sqrt{\pi}}\,,
\end{equation}
where $v_\mathrm{c}=(GM/a)^{1/2}$, $X=v_\mathrm{c}/(\sqrt{2}\,\sigma)$, and the cluster density $\rho(r)$ and one-dimensional velocity dispersion $\sigma(r)$ are measured directly from the initial snapshot. With the standard Coulomb logarithm $\ln\Lambda=\ln(M/m_\IMBH)\simeq7.6$ the prediction (dashed curve) matches the initial descent of the $\theta_\IMBH=90^\circ$ and $135^\circ$ models during the first ${\sim}2$~Myr, but over the full $10$~Myr the orbit of the IMBH decays more slowly by a factor of two. Thus, halving the Coulomb logarithm describes the measured tracks much better (dotted curve). The prograde $\theta_\IMBH=45^\circ$ model closely follows this prediction throughout its track, while the $\theta_\IMBH=90^\circ$, $135^\circ$ and $150^\circ$ models sink faster in the beginning but converge to it by $10$~Myr. Such a reduction of the dynamical friction relative to the standard estimate is within the uncertainties of the formula, which is sensitive to the choice of the maximum impact parameter and to the assumed background velocity distribution \citep{Just2011, AntoniniMerritt2012}.

While the lower-mass ($m_\IMBH=500\msun$) prograde model closely follows the modified $\ln\Lambda=\frac12\ln(M/m_\IMBH)$ 
prescription (pink solid and grey dotted curve), the retrograde model (light-green curve) migrates much more slowly. A plausible explanation is the different dynamical friction from the disc stars. The lower relative velocities of encounters with a prograde IMBH can strengthen this effect and faster retrograde encounters can weaken it.

Finally, the right panel of Fig.~\ref{fig:a_e_100k} shows the evolution of the eccentricities for the corresponding models. 
Most $m=2000\msun$ models circularise as their orbits decay; the $\theta_\IMBH=90^\circ$ model, for example, declines from $e_\IMBH=0.33$ to below $0.1$ by $10$~Myr. The exception is the $\theta_\IMBH=150^\circ$ model, whose eccentricity 
varies between $0.3$--$0.5$.
The initial rapid eccentricity growth to $e_\IMBH=0.5$ occurs during the fast initial migration episode of the IMBH, which counter-rotates with respect to the radially overlapping disc stars, consistent with the secular dynamical anti-friction models of \citet{Madigan2012}.
On the other hand, the IMBH with $\theta_\IMBH=135^\circ$ and $e_\IMBH=0.7$ (dark green curve) reduces its eccentricity and eventually varies between 0.4 and 0.2. The IMBH with $\theta_\IMBH=135^\circ$ and $e_\IMBH=0.33$ (light-green curve) varies its eccentricity stochastically between 0.1 and 0.4.

%#############
\section{Effects of the IMBH on the stellar disc}
\label{sec:disc-imbh}
%#############

We examine the disc thickness (Section~\ref{subsec:warp}), orbital-plane substructure (Section~\ref{subsec:substructure}) and eccentricities (Section~\ref{subsec:ecc-exc}). Throughout, `disc stars' denotes stars initially in the disc, irrespective of their subsequent orbits.

\subsection{Disc thickening and warping}
\label{subsec:warp}

\begin{figure*}
	\begin{centering}
		\includegraphics[width=\linewidth]{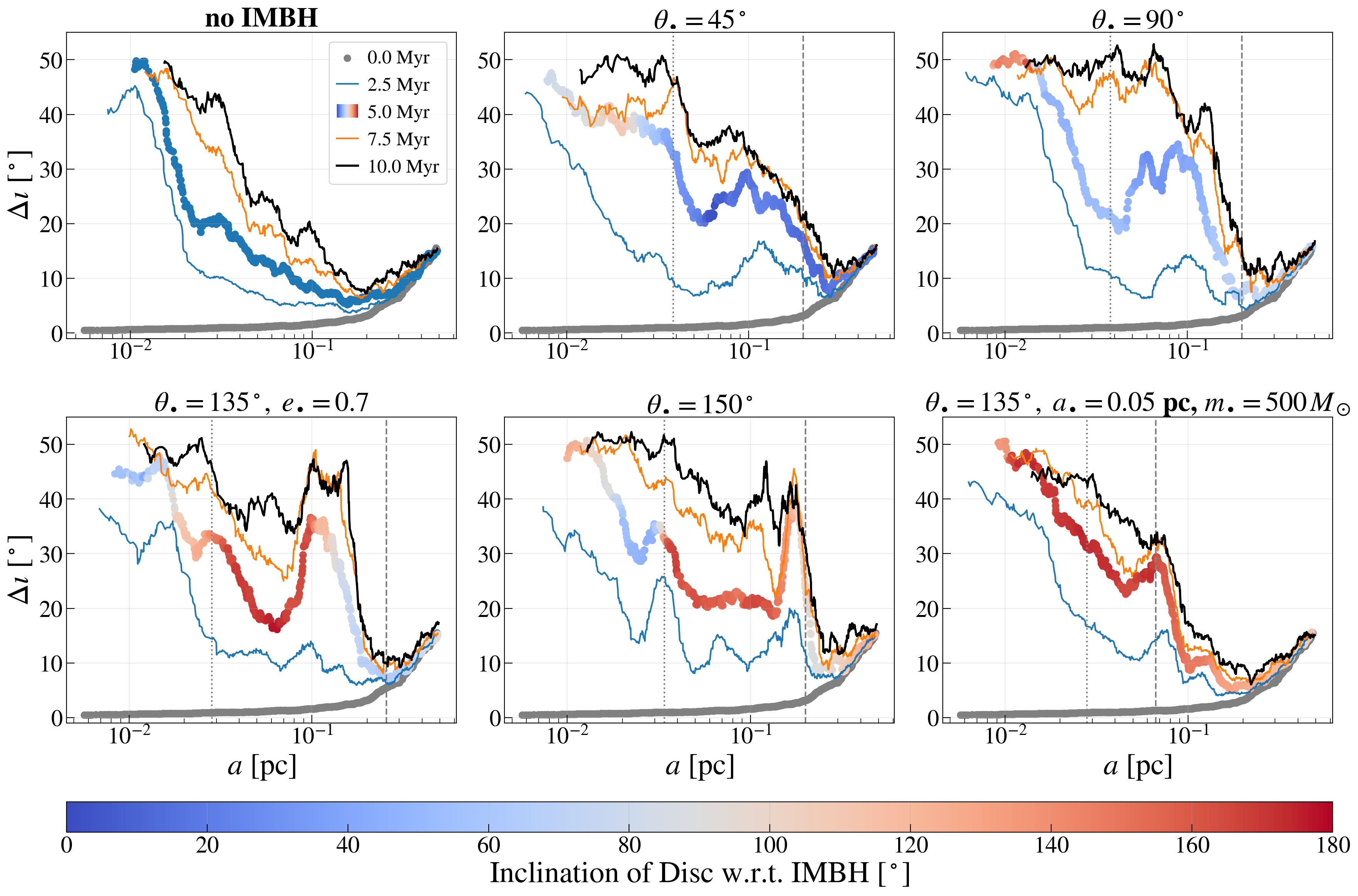}
		\par\end{centering}
	\caption{Thickness (warp) of the stellar disc, $\Delta\iota$, as a function of
	semi-major axis. Each panel corresponds to a model according to its title. Each panel shows five snapshots, at $t=0$, $2.5$, $5$, $7.5$ and $10$~Myr; the initial ($t=0$) disc is shown as grey points and the $t=10$~Myr state as a black line. The points at $t=5$~Myr are additionally colour-coded by the inclination of the local disc angular-momentum vector with respect to the IMBH, so that bluer (redder) colours denote regions rotating
	in the same (opposite) sense as the IMBH; the shared colour bar gives this inclination in degrees. In the no-IMBH panel, where no such reference direction exists, the $t=5$~Myr points are shown in plain blue. In the
	panels with an IMBH the vertical dashed and dotted lines mark the IMBH's initial apocentre and its pericentre at $10$~Myr, respectively.}
	\label{fig:L-space}
\end{figure*}

As in \PaperI, we quantify the local thickness of the disc by the root-mean-square warp angle $\Delta\iota=\arccos\sqrt{q}$, where $q$ is the principal eigenvalue of the quadrupole moment tensor of the angular momenta of the 50 stars nearest in semi-major axis; $\Delta\iota=0^\circ$ corresponds to a razor-thin disc and $\Delta\iota=54.7^\circ$ to an isotropic distribution. Figure~\ref{fig:L-space} shows $\Delta\iota$ as a function of semi-major axis for different models.

The disc thickens with time in all models, with $\Delta\iota$ in the inner disc reaching several tens of degrees by $10$~Myr. The colour coding at $t=5$~Myr shows that in the models with a retrograde IMBH ($\theta_\IMBH=135^\circ$, $e_\IMBH=0.7$ and $\theta_\IMBH=150^\circ$) the stars in the warped region are strongly inclined, and in places counter-rotating, with respect to the IMBH, whereas in the prograde ($\theta_\IMBH=45^\circ$) model they stay closer to alignment. 

The thickening in Fig.~\ref{fig:L-space} arises from two separate effects. First, the disc thickens even without an IMBH. In the no-IMBH panel, $\Delta\iota$ reaches $\simeq40^\circ$--$50^\circ$ in the inner disc and declines outward. In \PaperI this baseline thickening did not occur; here it is driven mostly by the stochastic torques exerted on the disc stars by the live sphere (vector resonant relaxation), and to a lesser extent by two-body
relaxation and scalar resonant relaxation, as quantified in \citet{Panamarev2022}. The resulting anticorrelation between disc thickness and semi-major axis, also found there, arises because the reorientation of the stellar orbits by the cluster proceeds more slowly at larger radii, where the resonant-relaxation timescale is longer. Second, the IMBH adds a localised signature on top of this relaxation-driven thickening, confined to the radial range it sweeps between the dotted (pericentre) and dashed (apocentre) lines. The outermost disc remains thin and coherent in every model, because there the cluster-induced reorientation is slowest and the disc self-torque dominates over the differential precession induced by the IMBH, consistent with the disruption criterion of \PaperI.

\subsection{Orbital-plane substructure}
\label{subsec:substructure}

The distribution of the disc stars' orbital angular-momentum directions on the sky is the quantity from which the existence of the young stellar disc in the Galactic centre was originally inferred \citep{LevinBeloborodov2003, Bartko2009, Yelda2014}. More recent observations reveal a more complex distribution, with several substructures in the projected map of the angular-momentum directions \citep{Ali2020, vonFellenberg2022, Siyao2023}. We therefore examine how the IMBH--disc interaction reshapes the Aitoff projection of these directions for a selection of the models of Section~\ref{sec:simulations}.

\begin{figure*}
    \centering
    \includegraphics[width=\linewidth]{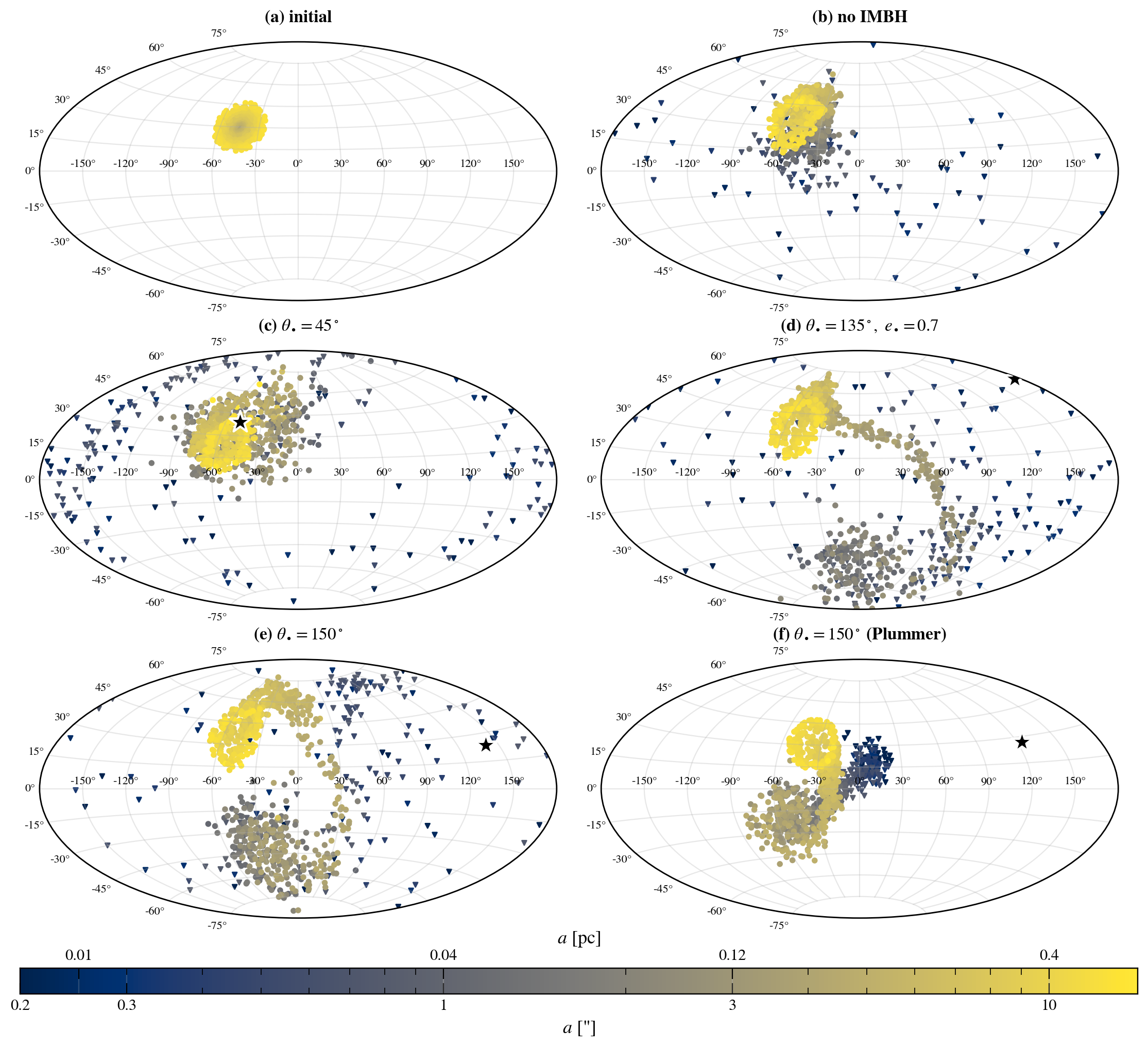}
    \caption{Directions of the disc stars' orbital angular-momentum vectors on an Aitoff (equal-area) projection, coloured by semi-major axis $a$; the black star marks the IMBH, and triangles denote the innermost stars ($a\le0.04$~pc). Panel (a) shows the initial disc, which is identical in all models; panel (b) the disc after $5$~Myr in the model without an IMBH; panels (c)--(e) the disc after $5$~Myr in three live-NSC models with a $2000\msun$ IMBH -- a prograde orbit ($\theta_\IMBH=45^\circ$, panel c), and two retrograde orbits ($\theta_\IMBH=135^\circ$, $e_\IMBH=0.7$, panel d; and $\theta_\IMBH=150^\circ$, panel e); and, for comparison, panel (f) the $\theta_\IMBH=150^\circ$, $2000\msun$ model with a smooth Plummer potential. We show detailed evolution in Appendix~\ref{app:skyev}.}
    \label{fig:skymaps}
\end{figure*}

Figure~\ref{fig:skymaps} contrasts the initial distribution of the orbital planes, shown in panel (a), identical in all models, with the distribution after $5$~Myr in a range of models; the corresponding $t=0$--$10$~Myr sequences for the individual live-sphere models are shown in Appendix~\ref{app:skyev}. Panel (b) shows the disc evolved in the live sphere without an IMBH, in which the inner and outer stars already acquire somewhat
different azimuthal orientations. In the retrograde $\theta_\IMBH=150^\circ$ model (panel e) the orbital planes separate into several concentrations at different semi-major axes, resembling the observed substructures noted above.

Panels (e) and (f) compare the live-cluster and fixed-Plummer models. The same model at the same epoch also begins to fragment in the fixed Plummer background, but there the components remain close together on the sky and
the innermost stars stay confined near the disc. In the live sphere the components attain far larger mutual separations -- comparable to the separations between the observed structures -- and the innermost stars
(triangles) disperse across the whole sky. In \PaperI such separations developed only after $10$--$20$~Myr. The stochastic torques of the live cluster, aided by the inward migration of the IMBH
(Fig.~\ref{fig:a_e_100k}), accelerate the process by a factor of a few. This confirms the expectation of \PaperI that the timescales obtained with a static background are upper limits. In the live cluster, such separations develop
by ${\sim}5$~Myr, consistent with the estimated age of the young stellar population in the Galactic centre ($6\pm2$~Myr, \citealt{PaumardEtAl2006}; $2.5$--$5.8$~Myr, \citealt{Lu2013}).

To characterise these substructures, we apply a two-stage clustering to the disc stars of the retrograde model ($\theta_\IMBH=150^\circ$, $a_\IMBH=0.15$~pc, $e_\IMBH=0.33$, $m_\IMBH=2000\msun$) at $t=5$~Myr, in the combined space of angular-momentum direction and semi-major axis. The first stage uses \textsc{hdbscan} \citep{Campello2013, McInnes2017}, a hierarchical density-based algorithm that identifies clusters as regions of high local density separated by sparser regions and leaves stars outside any such region unclustered. In the second stage, the outermost component is subdivided by single-linkage agglomerative clustering -- a hierarchical procedure that repeatedly merges the two groups with the smallest distance between their closest members.

\begin{figure*}
    \centering
    \includegraphics[width=\linewidth]{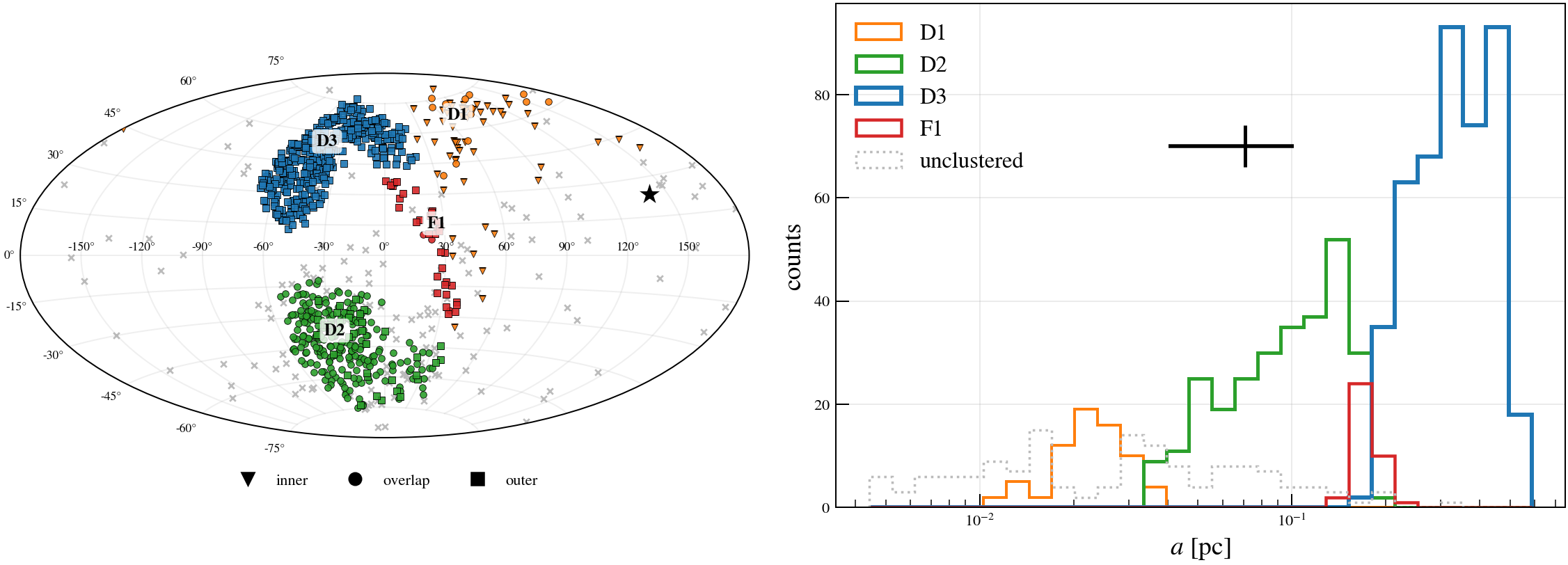}
    \caption{Substructure of the disc in the retrograde model ($\theta_\IMBH=150^\circ$, $a_\IMBH=0.15$~pc, $e_\IMBH=0.33$, $m_\IMBH=2000\msun$) at $t=5$~Myr. \emph{Left:} directions of the stellar orbital angular-momentum vectors on an Aitoff projection, coloured by the component to which each star is assigned in Section~\ref{subsec:substructure}, namely three disc-like structures (D1, D2, D3) and a filament (F1).
    Stars that do not belong to any of the structures are shown in grey. The marker shapes indicate the radial positions of the stars with respect to the IMBH: triangles -- inner stars; circles -- overlapping stars; and squares -- outer stars, as defined in Eq.~\eqref{eq:orb-cond}. The black star symbol marks the IMBH. \emph{Right:} distribution of the semi-major axes of the stars in each component; the horizontal bar marks the orbital span of the IMBH and the tick its semi-major axis.}
    \label{fig:composite_sky}
\end{figure*}

Figure~\ref{fig:composite_sky} shows the result. The clustering identifies four components: three disc-like structures -- D1 (innermost), D2 (middle), and D3 (outermost) -- together with a filament F1 split off from the outer disc. The median semi-major axis of F1 falls between those of D2 and D3; the remaining stars are left unclustered. The middle component D2 -- dominated by the stars radially overlapping the IMBH orbit --
is strongly misaligned with the outer disc D3. This decomposition closely resembles the structure identified in \PaperI for the smooth spherical background models -- three disc-like components and a filament, labelled in the same way -- and is already in place at $t=5$~Myr. Some stars remain outside the identified substructures, consistent with the observations \citep{Yelda2014, Siyao2023}.

Overall, the components behave as intact sub-units of the initial disc, as expected in the intact-fragment regime of the disruption criterion of \PaperI, at least up to $6$~Myr. At later times, the inner region becomes dispersed due to relaxation in the background NSC. We show Aitoff projections of these stars at different snapshots for the corresponding model, as well as for the other models, in Appendix~\ref{app:skyev}.

\begin{figure*}
    \centering
    \includegraphics[width=\linewidth]{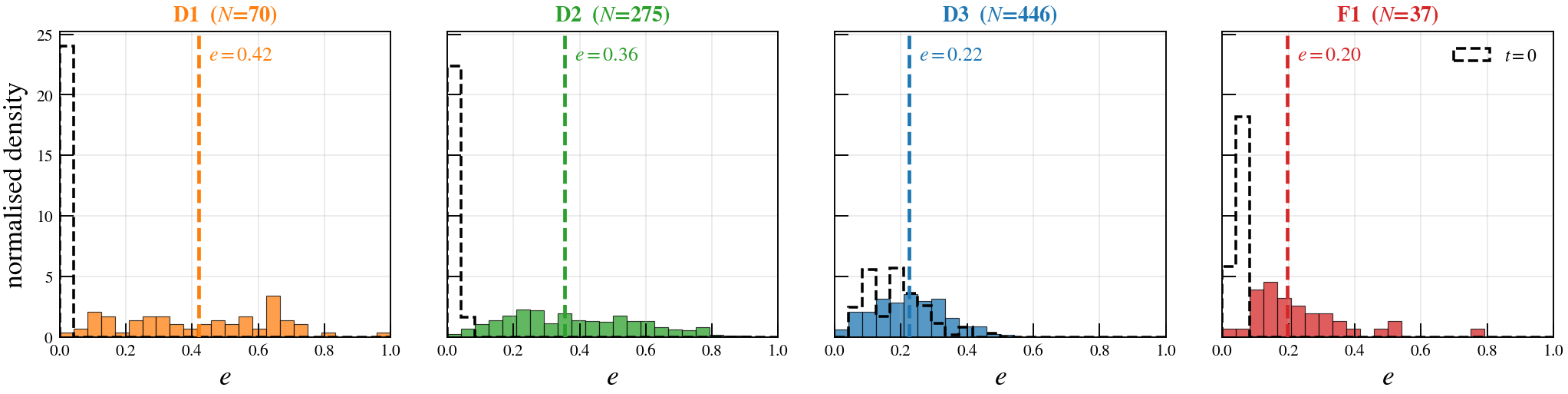}
    \caption{Eccentricity distributions of the four disc components of Fig.~\ref{fig:composite_sky} (D1, D2, D3, F1 from left to right) in the same model at $t=5$~Myr as in Fig.~\ref{fig:composite_sky}. In each panel the dashed black histogram shows the initial eccentricities of the same stars. The vertical dashed line marks the median eccentricity at $t=5$~Myr, whose value is annotated. Each distribution is normalised to unit area.}
    \label{fig:composite_ae}
\end{figure*}

\subsection{Eccentricity excitation}
\label{subsec:ecc-exc}

The eccentricity distribution of young stars could also be a diagnostic of a hidden IMBH in a nuclear star cluster. First note that in the absence of an IMBH, starting from a cold stellar disc, two-body relaxation generally reduces the random velocities and eccentricities of heavier stars, while lighter stars are heated \citep{Alexander+2007,Mikhaloff_Perets2017}. Scalar resonant relaxation also produces a mass-dependent eccentricity distribution \citep{Fouvry2018,Gruzinov2020}. Our previous $N$-body simulations showed that these effects are suppressed if the disc is embedded in a live spherical cluster \citep{Panamarev2022}. The eccentricity of an IMBH in the cluster, if present, strongly decreases in the presence of a prograde disc and evolves more slowly in a retrograde configuration (\citealt{Szolgyen2021}; \PaperI). An IMBH is expected to have an impact primarily on the radial distribution of the stellar eccentricities mostly independently of stellar mass. The IMBH strongly excites eccentricities specifically in the radially overlapping population, while the outer population is largely unchanged and the inner population is affected less strongly (\PaperI). For the prograde configurations, the radially overlapping population even develops a high-$e$ tail reaching $e\gtrsim0.7$--$0.9$, whereas the retrograde cases show a broader but less shifted distribution. Here we examine how a live spherical cluster modifies these conclusions.

Figure~\ref{fig:composite_ae} shows the eccentricity distribution of each component, comparing the state at $t=5$~Myr with the initial ($t=0$) eccentricities of the same stars. The inner components D1 and D2 reach the broadest, highest-eccentricity distributions (median $e\simeq0.42$ and $0.36$), whereas the outer disc D3 and the
filament F1 remain closer to their initially low eccentricities (median $e\simeq0.2$). The moderate median eccentricities of D1 and D2 are comparable to those inferred for the observed clockwise-disc substructures (\citealt{vonFellenberg2022}; $e\sim0.3$--$0.4$). The angular momenta of D1 and D2 are misaligned by $131^\circ$, close to the ${\simeq}130^\circ$ between the angular-momentum directions of the inner clockwise disc (CW1) and the counter-clockwise structure (CCW/F1) identified by \citet{vonFellenberg2022}. The outer disc D3 keeps eccentricities close to its initial values (median $e=0.19$, $0.22$ and $0.24$ at $0$, $5$ and $10$~Myr), somewhat lower than the median of the observed outer clockwise disc ($e\simeq0.4$; \citealt{vonFellenberg2022}), although it lies within its interquartile range. 

\begin{figure*}
    \centering
    \includegraphics[width=\linewidth]{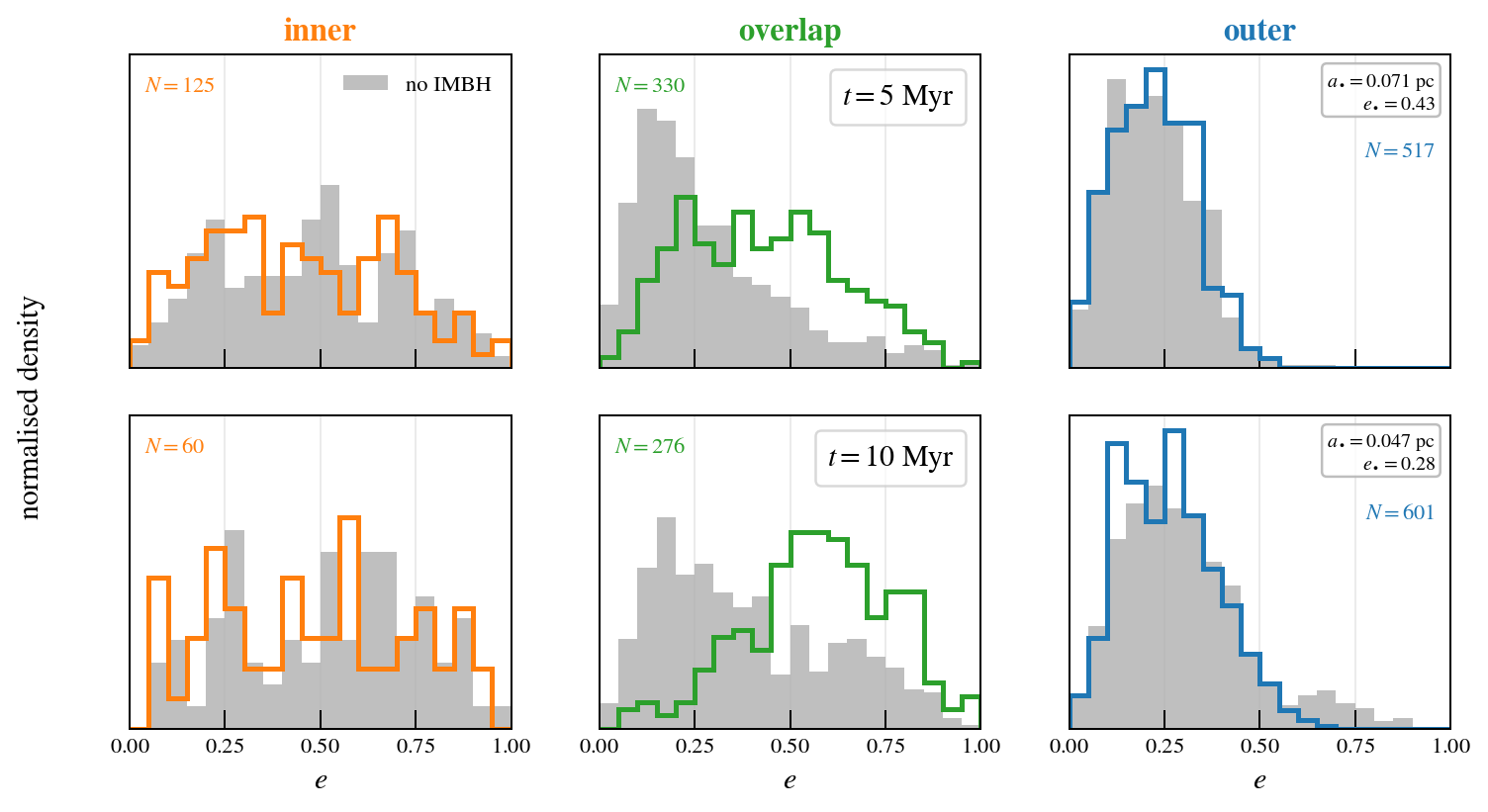}
    \caption{Eccentricity distributions of the disc stars in the retrograde model ($\theta_\IMBH=150^\circ$, $a_\IMBH=0.15$~pc, $e_\IMBH=0.33$, $m_\IMBH=2000\msun$; coloured lines), separated into the stars orbiting
    entirely inside the IMBH's instantaneous pericentre (left column), radially overlapping with its orbit (middle column) and entirely outside its apocentre (right column), at $t=5$~Myr (top row) and $t=10$~Myr (bottom row). The grey filled histograms show the model without an IMBH, restricted to the same ranges of semi-major axis. All distributions are normalised to unit area; the number of stars in each regime and the IMBH's instantaneous semi-major axis and eccentricity are annotated.}
    \label{fig:ecc_regimes}
\end{figure*}

Figure~\ref{fig:ecc_regimes} isolates the part of the eccentricity excitation driven by the IMBH. It compares, for the fiducial retrograde model, the eccentricity distributions of the disc stars in the three dynamical regimes defined by the IMBH's instantaneous orbit -- inner, radially overlapping and outer -- with those in the model without an IMBH evolved in the same cluster, at $t=5$ and $10$~Myr. In the radially overlapping region, the IMBH drives the distribution to systematically higher eccentricities than in the no-IMBH baseline. In the inner region, the two distributions are similar, because the heating there is dominated by relaxation against the cluster, which broadens the baseline itself. The outer region is only mildly heated, and equally so in both models. This contrasts with \PaperI, where the smooth static background produced no baseline eccentricity heating, so nearly all of the excitation was attributable to the IMBH. In the live NSC, the IMBH-specific signature is confined to the radial range swept by its orbit. By $t=5$~Myr, the IMBH has raised the eccentricities of the disc stars within this radial range to the moderate values inferred for the observed clockwise-disc substructures ($e\sim0.3$--$0.4$), whereas in the inner disc comparable eccentricities arise from relaxation in the live NSC alone.

%#############
\section{The innermost stars}
\label{sec:inner}
%#############

We compare the orbital parameters of the innermost disc stars, evolved with and without an IMBH in the live NSC, with those of the observed S-stars. Throughout, we select the disc stars with semi-major axis $a\le0.04$~pc, which at the assumed Galactic-centre distance of $R_0=8$~kpc corresponds to an angular scale of $1''$ and to the region
occupied by the observed S-cluster \citep[e.g.][]{Ghez2008, Gillessen2009, Gillessen2017}. Since the observed orbits are known in three dimensions, we compare the Keplerian semi-major axes of the simulated stars with the semi-major axes obtained from the observed stars' fitted three-dimensional orbital solutions.

\subsection{The observed S-star sample}
\label{subsec:sstar-obs}

The census of young stars in the inner arcsecond is assembled from several observational programmes that differ in instrument, selection and depth. We consider four catalogues. The catalogue of \citet{Gillessen2017}, based on more than two decades of \textsc{naco} and \textsc{sharp} adaptive-optics imaging and spectroscopy, provides the reference set of stellar orbits in the central arcsecond. The sample of \citet{vonFellenberg2022} is based on \textsc{sinfoni} spectroscopy, with astrometry taken from the adaptive-optics monitoring programme, and is restricted to stars spectroscopically confirmed as young through their Br$\gamma$ line. For stars in common with \citet{Gillessen2017}, \citet{vonFellenberg2022} update the earlier fits, extending the 1992--2016 monitoring baseline by a further ${\sim}5$~yr of astrometry -- including \textsc{gravity} interferometric positions -- together with re-reduced \textsc{sinfoni} radial velocities. Where the two catalogues disagree, we therefore adopt the more recent \citet{vonFellenberg2022} solutions. The differences are substantial for the short-period star S18, whose orbit remains weakly constrained ($e=0.47$ in \citealt{Gillessen2017} versus $0.24$ in \citealt{vonFellenberg2022}), for S29 ($e=0.73$ versus $0.97$) and for S42 ($e=0.57$ versus $0.75$, and $a=0.95''$ versus $0.38''$). For the near-radial S14, the two eccentricities agree within the uncertainties ($e=0.976$ versus $0.99$), but at $e\simeq0.98$ even this small shift halves the pericentre $r_\mathrm{p}=a(1-e)$, to which the zone-of-avoidance classification of Section~\ref{subsec:zoa} below is sensitive.

\begin{figure}
\centering
\includegraphics[width=\columnwidth]{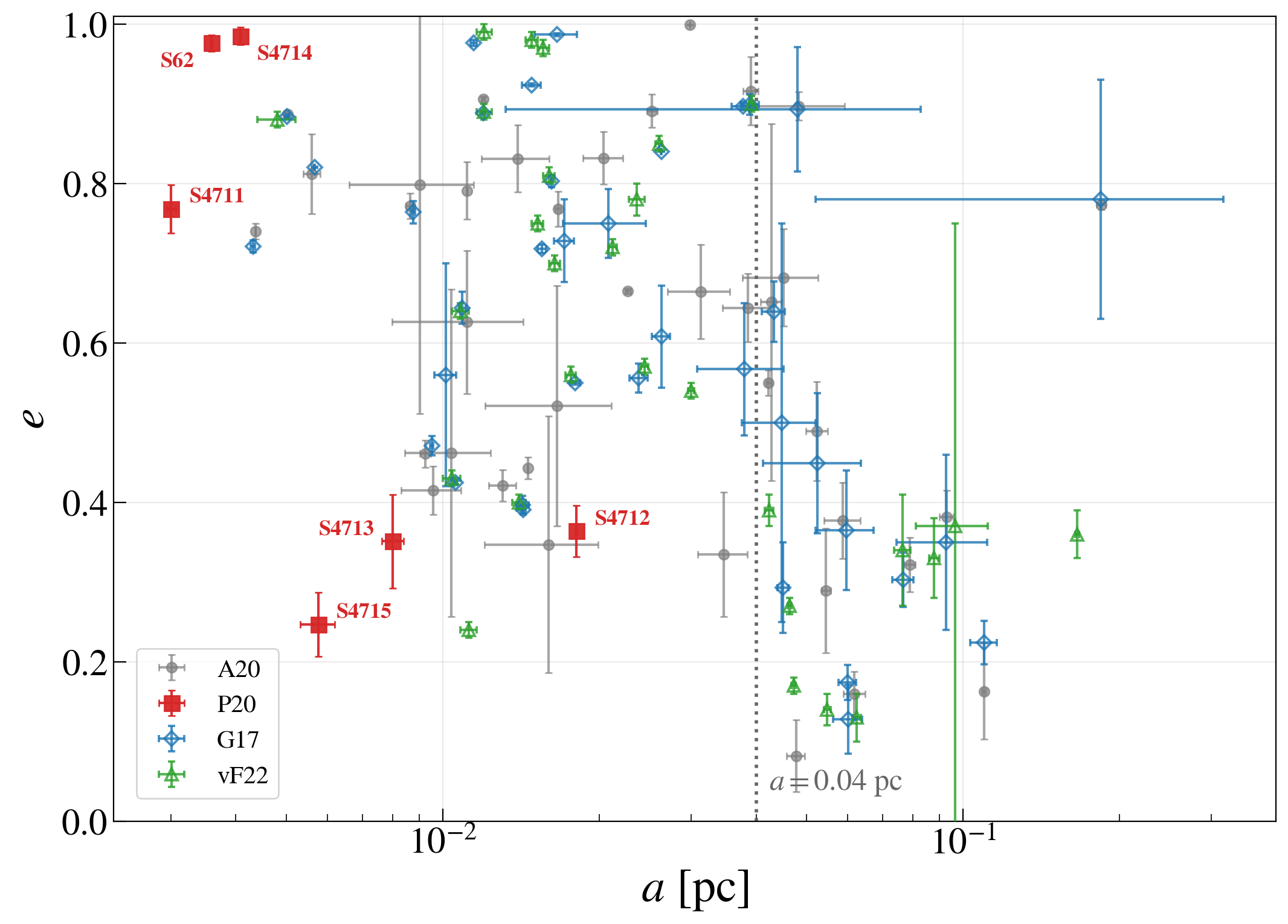}
\caption{Eccentricity versus semi-major axis for the observed stars in the
Galactic centre. Points show the catalogues of \citet{Gillessen2017} (G17; 37 stars), \citet{vonFellenberg2022} (vF22; 28) and \citet{Ali2020} (A20; 38), with error bars where available; the six fast-moving stars reported by \citet{Peissker2020} (P20) are highlighted as separate labelled symbols. The catalogues partly overlap.
The vertical line marks the $a=0.04$~pc ($1''$) boundary used to define the innermost sample.}
\label{fig:sstars_ea}
\end{figure}

\begin{figure*}
\centering
\includegraphics[width=\linewidth]{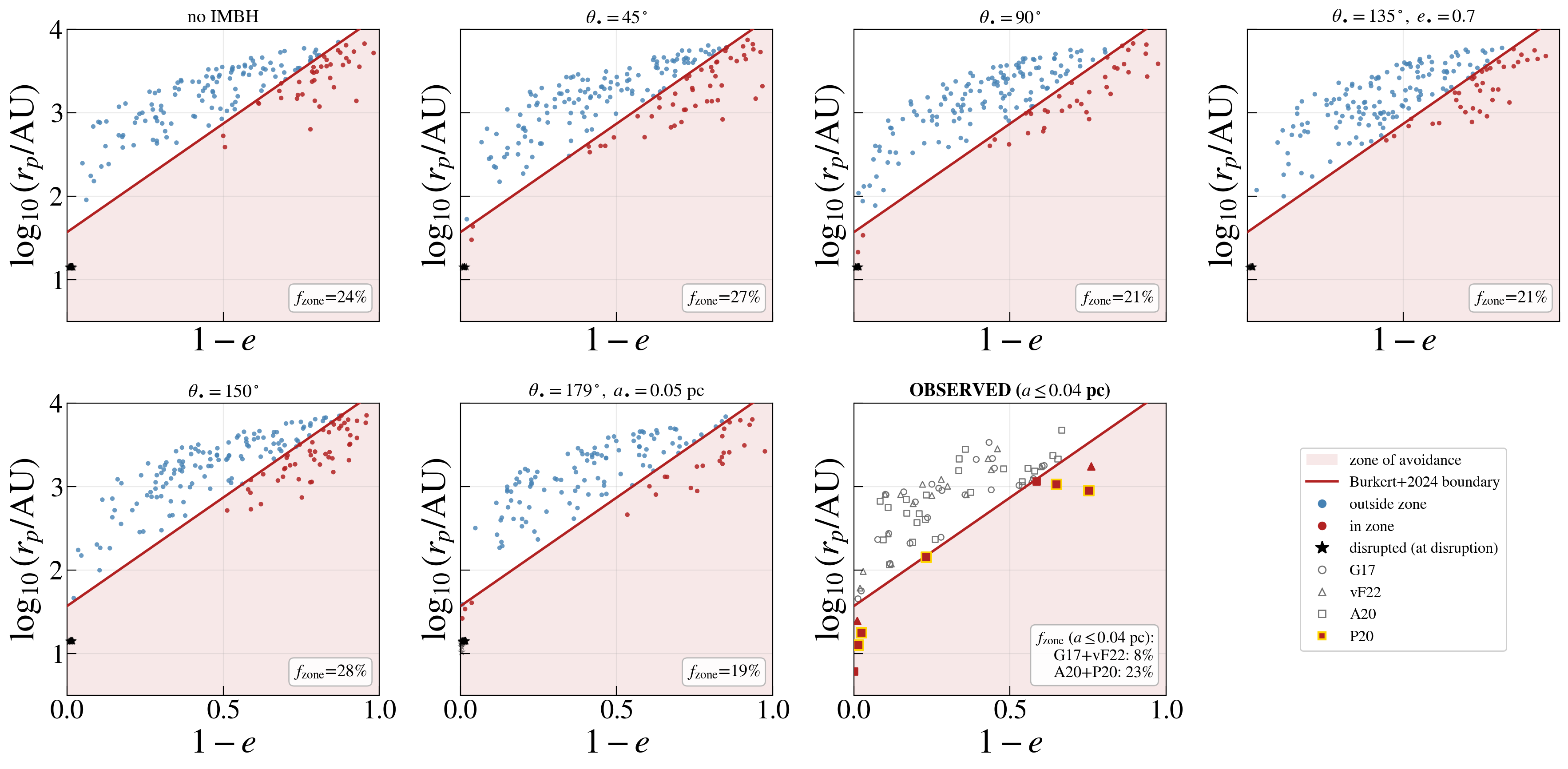}
\caption{Innermost stars ($a\le0.04$~pc) in the pericentre--eccentricity plane at $t=5$~Myr, for the model without an IMBH, the $2000\msun$ IMBH models, and the observed sample (last panel). In each panel, the solid line and shaded region show the zone-of-avoidance boundary of Eq.~\eqref{eq:zoa} \citep{Burkert2024}; stars are coloured according to whether they are within the zone (red) or not (blue; open grey symbols in the observed panel), and the in-zone fraction $f_\mathrm{zone}$ is annotated. Black star symbols mark each model's disc stars tidally disrupted by the SMBH before the epoch shown, using their orbital elements at the moment of disruption; they are not counted in $f_\mathrm{zone}$. In the observed panel, the \citet{Gillessen2017} (G17), \citet{vonFellenberg2022} (vF22), and \citet{Ali2020} (A20) catalogues are shown with different symbols, while the \citet{Peissker2020} (P20) stars are highlighted; the annotated in-zone fractions refer to the combined G17$+$vF22 and A20$+$P20 samples. }
\label{fig:zone_panels}
\end{figure*}

\begin{figure*}
\centering
\includegraphics[width=\linewidth]{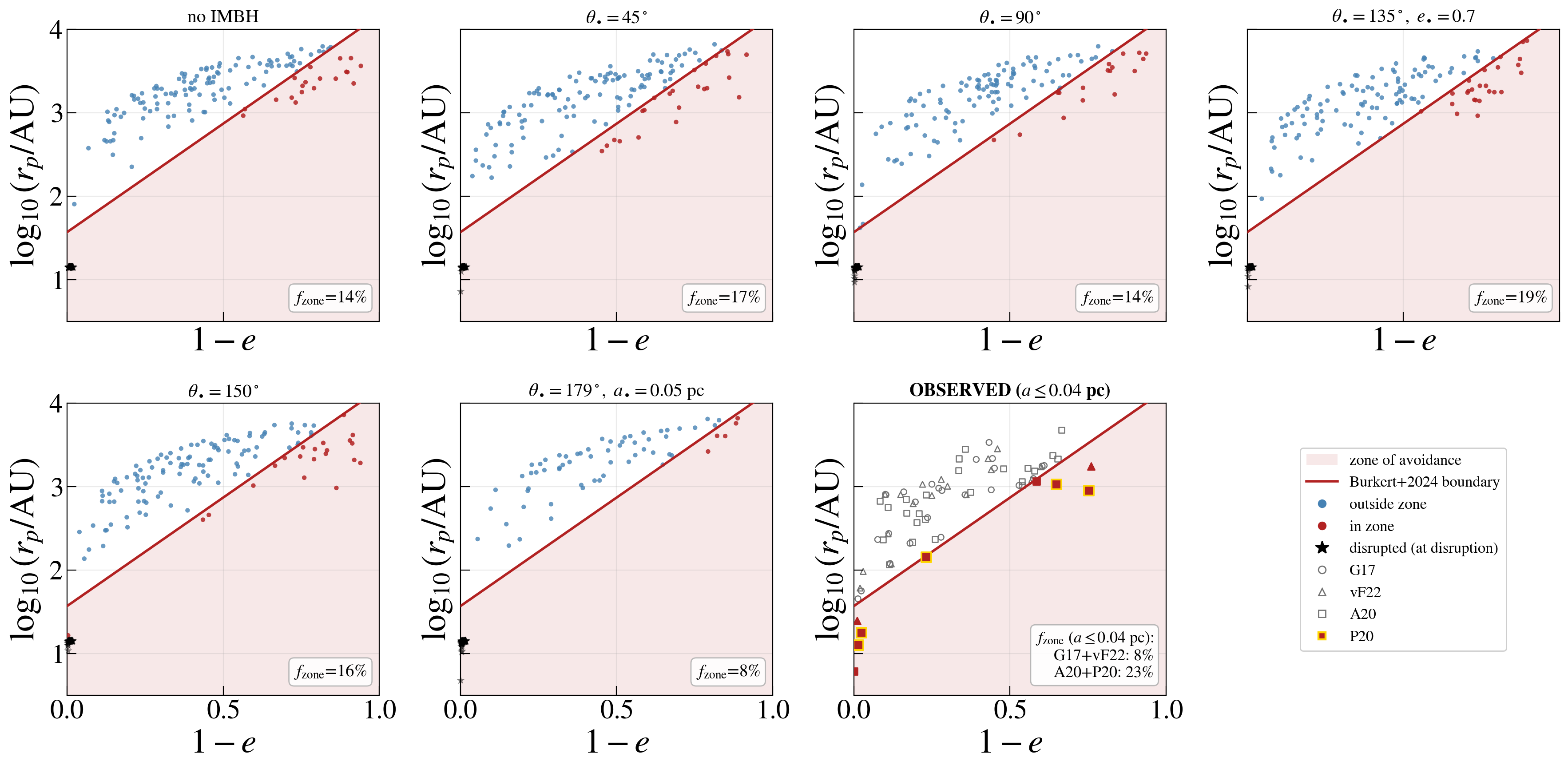}
\caption{As Fig.~\ref{fig:zone_panels}, but at $t=10$~Myr.}
\label{fig:zone_panels_10}
\end{figure*}

The catalogue of \citet{Ali2020} is based on \textsc{naco} $K_\mathrm{s}$-band imaging astrometry of the central arcsecond. Its sample consists of B-type stars with $K_\mathrm{s}<18$ detected in every epoch (i.e.\ in each yearly imaging data set) of the 2002--2019 \textsc{naco} monitoring, with radial velocities and spectral classifications taken from the literature. Finally, \citet{Peissker2020} report a set of faint, fast-moving stars (S62 and S4711--S4715) recovered from high-pass-filtered \textsc{naco} imaging. Several of these have not been confirmed by subsequent \textsc{gravity} observations \citep{Gravity2021, GravityDeep2022}, so we treat them as a separate addition to the imaging catalogues. Figure~\ref{fig:sstars_ea} shows eccentricity as a function of semi-major axis for these catalogues. Since the catalogues partly overlap, stars common to several of them appear once for each catalogue in the scatter plots (Figs.~\ref{fig:sstars_ea}, \ref{fig:zone_panels} and \ref{fig:zone_panels_10}). In the combined \citet{Gillessen2017} and \citet{vonFellenberg2022} sample each star is counted once, with the \citet{vonFellenberg2022} solution adopted.

\subsection{Comparison with the observed S-stars}
\label{subsec:sstar-sim}

In the comparisons below, we assume that the S-stars formed in the same star-formation episode as the disc stars, and define the S-star analogues in the simulation as the disc stars whose semi-major axes satisfy $a\le0.04$~pc at each snapshot. Their membership is therefore re-evaluated with time.

\begin{figure*}
\centering
\includegraphics[width=\linewidth]{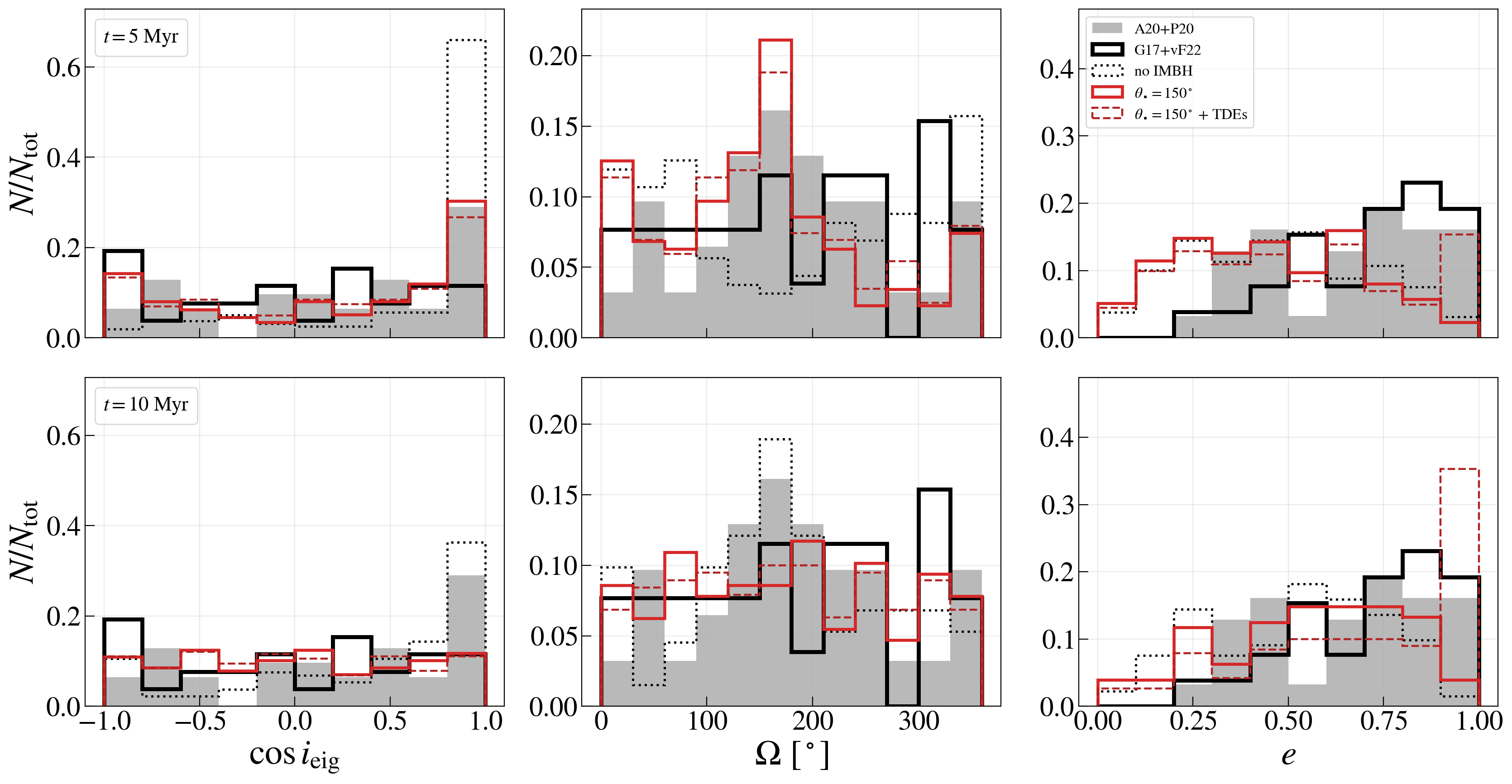}
\caption{Normalised distributions of the orbital elements of the innermost stars ($a\le0.04$~pc) for the fiducial retrograde model ($\theta_\IMBH=150^\circ$, $a_\IMBH=0.15$~pc, $e_\IMBH=0.33$, $m_\IMBH=2000\msun$), compared with the observed S-stars. Columns show the cosine of inclination $\cos i_\mathrm{eig}$ (left), the longitude of the ascending node $\Omega$ (middle) and the eccentricity $e$ (right); the top and bottom rows correspond to $t=5$ and $10$~Myr. In each panel the shaded histogram is the A20$+$P20 imaging sample (\citealt{Ali2020} $+$ \citealt{Peissker2020}), the thick black line the combined G17$+$vF22 sample (\citealt{Gillessen2017} $+$ \citealt{vonFellenberg2022}; Section~\ref{subsec:sstar-obs}), the solid red line is the simulated inner disc, the thin dashed red line adds to it the disc stars tidally disrupted by the central SMBH during the run, at their disruption-epoch orbital elements, and the dotted black line is the inner disc of the model without an IMBH. Each histogram is normalised to unit sum. The node $\Omega$ is shown after a single, snapshot-independent constant shift, common to both models, that aligns the simulated peak with the observed one, since its zero-point is
arbitrary.}
\label{fig:sstars_hist}
\end{figure*}

For the fiducial retrograde model, Figure~\ref{fig:sstars_hist} compares the orbital-element distributions of this simulated population with those of the observed sample. The panels compare the cosine of the orbital inclination $\cos i_\mathrm{eig}$ (measured relative to the principal eigenvector of the inner region), the longitude of the ascending node $\Omega$, and the eccentricity $e$, at $t=5$ and $10$~Myr. Since the zero-point of the node $\Omega$ is arbitrary in the simulation frame, we apply a single constant offset -- the same for both snapshots -- so that the peak of the simulated node distribution coincides with the observed one.

In Figure~\ref{fig:sstars_hist}, the simulated S-stars (red) cover a broad range of $\cos i_\mathrm{eig}$ and eccentricity that overlaps the observed distributions. Their eccentricity distribution broadens between $t=5$ and $t=10$~Myr, and adding the tidally disrupted stars extends it further towards high $e$. After the arbitrary shift, the node distribution $\Omega$ is comparably concentrated in the simulation and in the observed catalogues.

These comparisons indicate that an initially cold inner disc is transformed, within the lifetime of the observed disc, into a population whose orbital elements span the ranges occupied by the S-stars. At $t=5$~Myr, the surviving population still retains some memory of its origin, as its inclination distribution peaks at $\cos i_\mathrm{eig}\simeq1$, like that of the imaging sample \citep{Ali2020, Peissker2020}. By $t=10$~Myr it flattens towards the broader distribution of the combined \citet{Gillessen2017} and \citet{vonFellenberg2022} sample. The model without an IMBH evolves more slowly, and its inclination distribution remains more sharply peaked at $\cos i_\mathrm{eig}\simeq1$ at both epochs, indicating that the IMBH accelerates the dispersal of the innermost orbital planes. Appendix~\ref{app:disc-relaxation} quantifies the changes in the disc stars' energies and angular momenta for this model.

In eccentricity, the combined \citet{Gillessen2017} and \citet{vonFellenberg2022} distribution peaks near $e\simeq0.8$. The inner stars of the fiducial model overlap this range, but their median eccentricity reaches only $e\simeq0.58$ at $10$~Myr, close to $e\simeq0.55$ in the no-IMBH control. The IMBH therefore raises the median only marginally, and both medians remain below the observed peak. The extreme near-radial tail ($e>0.98$) is reached by the disc stars at the moment of their tidal disruption, just before they are removed from the simulation (Section~\ref{sec:tde}). A disc origin can therefore account for a substantial part of the observed orbital-element distributions on the relevant timescale.

The disrupted stars also point to the Hills mechanism \citep{Hills1988} as a channel for the most eccentric S-stars. If some disc stars are binaries, the SMBH separates those that reach their tidal separation radius (Section~\ref{subsec:tde-binaries}). A separation can eject one component and leave the other on a tighter, highly eccentric orbit. Subsequent relaxation would broaden the initially sharp $e\simeq1$ distribution \citep{Generozov2020}. The estimates in Section~\ref{subsec:tde-hills} describe the number and eccentricities of the captured stars at separation. Recent observations indicate that some disc stars are indeed in binaries, providing a lower limit on the binary fraction of the young stars \citep{Gautam2024}.

\subsection{Zone of avoidance}
\label{subsec:zoa}

The observed S-stars tend to avoid a well-defined region of the plane spanned by the orbital pericentre distance $r_\mathrm{p}=a(1-e)$ and the eccentricity. Few stars are observed with both a small pericentre and a low eccentricity, and the \citet{Gillessen2017} catalogue contains none. This sparsely populated region is referred to as the `zone of avoidance' \citep{Burkert2024, Generozov2025}. Following \citet{Burkert2024}, we delineate it by the boundary
\begin{equation}\label{eq:zoa} \log_{10}\!\left(\frac{r_\mathrm{p}} {\mathrm{AU}}\right) < 1.57 + 2.6\,(1-e)\,.
\end{equation} We apply the same criterion to the simulated innermost disc stars ($a\le0.04$~pc) and to the observed catalogues.

Figure~\ref{fig:zone_panels} shows the innermost stars in the plane of $\log_{10}(r_\mathrm{p}/\mathrm{AU})$ versus $1-e$ at $t=5$~Myr, for the model without an IMBH, the five $2000\msun$ IMBH models of Table~\ref{tab:runs}, and the observed sample. Each model panel also marks the disc stars tidally disrupted before that epoch, using their orbital elements at the moment of disruption. Their distribution peaks at near-radial eccentricities ($e>0.98$), well inside the zone. Figure~\ref{fig:zone_panels_10} repeats the same construction at $t=10$~Myr.

\begin{figure}
\centering
\includegraphics[width=\columnwidth]{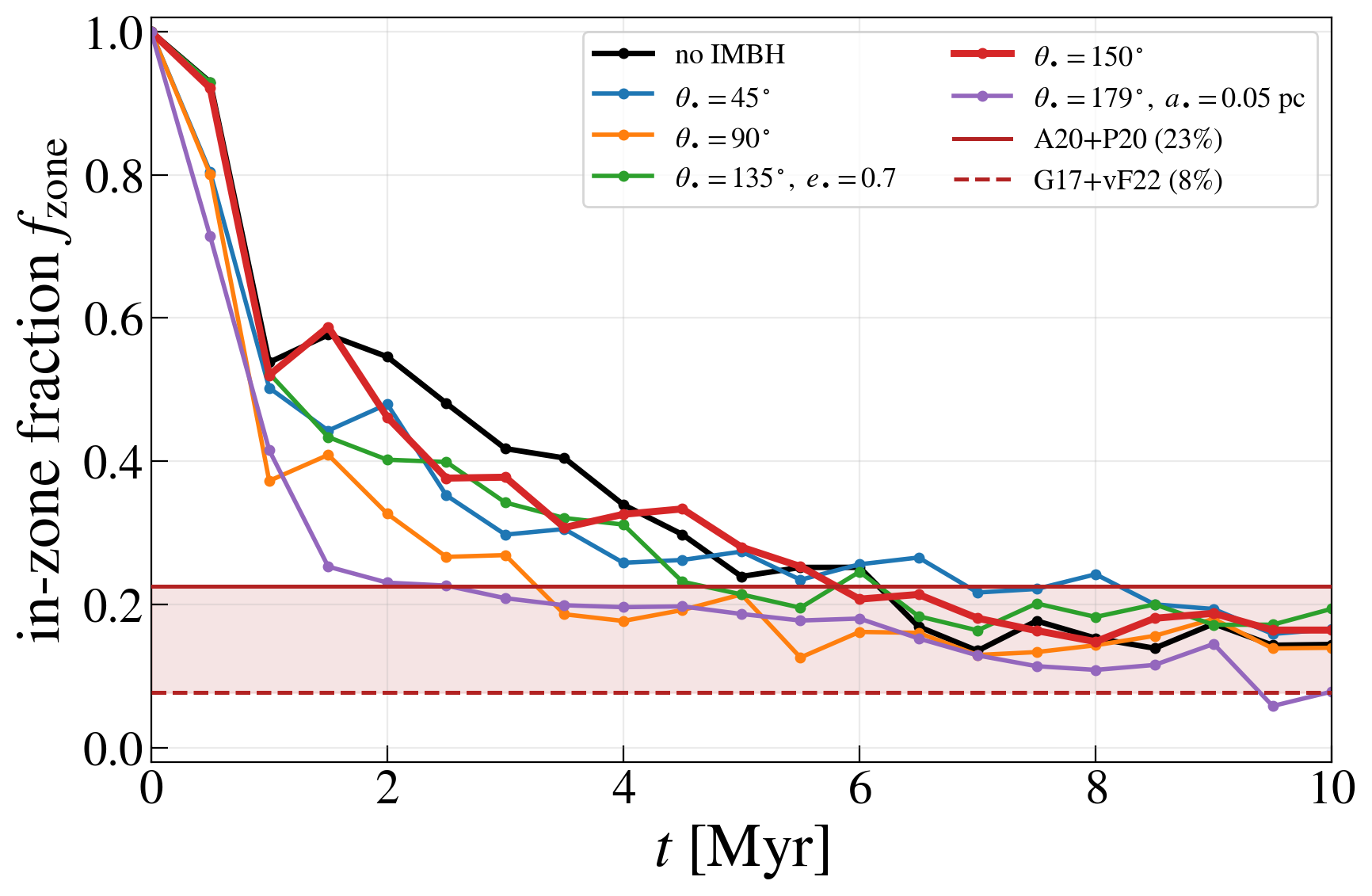}
\caption{Fraction of the innermost stars ($a\le0.04$~pc) lying inside the zone of avoidance, $f_\mathrm{zone}$, as a function of time for each simulation shown (one line per model, coloured as elsewhere in the paper). The shaded horizontal band spans the range of in-zone fractions of the combined observed samples, from G17$+$vF22 (\citealt{Gillessen2017} $+$ \citealt{vonFellenberg2022}; $8$\%; dashed edge) to A20$+$P20 (\citealt{Ali2020} $+$ \citealt{Peissker2020}; $23$\%; solid edge). See Section~\ref{subsec:sstar-obs} for the treatment of stars common to the G17 and vF22 catalogues.}
\label{fig:zone_time}
\end{figure}

Figure~\ref{fig:zone_time} shows the fraction of innermost stars inside the zone of avoidance, $f_\mathrm{zone}$, as a function of time for the same models. All runs start with every innermost star inside the zone, because the disc is initially cold and its stars have small pericentres at low eccentricity. As the disc evolves, $f_\mathrm{zone}$ decreases. By $10$~Myr it lies between $8$ and $19$\% in all models shown, within the range spanned by the observed samples (the shaded band), which extends from $8$\% for the combined \citet{Gillessen2017} and \citet{vonFellenberg2022} sample to $23$\% for the \citet{Ali2020} and \citet{Peissker2020} imaging sample. Even without an IMBH, the live NSC depletes the zone within $10$~Myr, with $f_\mathrm{zone}$ declining from $100$\% to ${\simeq}24$\% at 5~Myr and ${\simeq}14$\% at 10~Myr. An IMBH can accelerate this evolution. The $\theta_\IMBH=179^\circ$, $a_\IMBH=0.05$~pc model, whose pericentre is ${\simeq}0.03$~pc, reaches ${\simeq}8$\% by 10~Myr, whereas the models with $a_\IMBH=0.15$~pc reach values similar to that of the model without an IMBH by 10~Myr.

The comparison is limited by the small number of stars in both the simulations and the observed samples. Each model retains $120$--$180$ innermost stars at $5$--$10$~Myr ($64$ in the $\theta_\IMBH=179^\circ$ model at $10$~Myr), four to seven times more than the observed samples. The combined \citet{Gillessen2017} and \citet{vonFellenberg2022} sample contains $26$ stars with $a\le0.04$~pc, of which $8$\% (S14 and S18) lie in the zone; using the \citet{Gillessen2017} solutions for these two stars, the fraction would be zero. The recently discovered star S301 ($a=83$~mas, $e=0.98$; \citealt{Gravity2026}) also lies inside the zone of avoidance; its spectral type is not yet known, and if it is an early-type star, it raises the in-zone fraction of this sample to $11$\%. The imaging sample contains $31$ stars, of which $23\pm8$\% lie in the zone.

Both samples also contain a few late-type stars \citep{Gillessen2017}, none of which lies in the zone. Several stars, including the in-zone stars reported by \citet{Peissker2020}, have no confirmed spectral classification. Counting only spectroscopically confirmed early-type stars, both samples give in-zone fractions of ${\simeq}10$\%.

The in-zone fraction of the combined \citet{Gillessen2017} and \citet{vonFellenberg2022} sample depends on the adopted orbital solutions of only two stars, and five of the seven in-zone stars of the imaging sample are stars reported by \citet{Peissker2020}, several of which have not been confirmed (Section~\ref{subsec:sstar-obs}). Without these five stars, the imaging sample also gives ${\simeq}8$\%. The simulated in-zone fractions therefore lie within the range spanned by the combined observed samples.
The small observed fraction of stars in the zone of avoidance is therefore consistent with a disc origin of the innermost stars followed by relaxation, both with and without an IMBH of the masses and orbits considered in our study.

\citet{Zheng2026} also reproduce the observed zone of avoidance, at an age of $9$~Myr, and attribute it to the formation of the young stars in a common gaseous disc at $a\gtrsim0.01$~pc. In our simulations, every innermost star starts inside the zone, and relaxation against the NSC depletes it within ${\sim}10$~Myr, with or without an IMBH.

%###########################################################################
\section{Tidal disruptions and implications for the Hills mechanism}
\label{sec:tde}
%###########################################################################

\begin{figure*}
\centering
\includegraphics[width=0.8\linewidth]{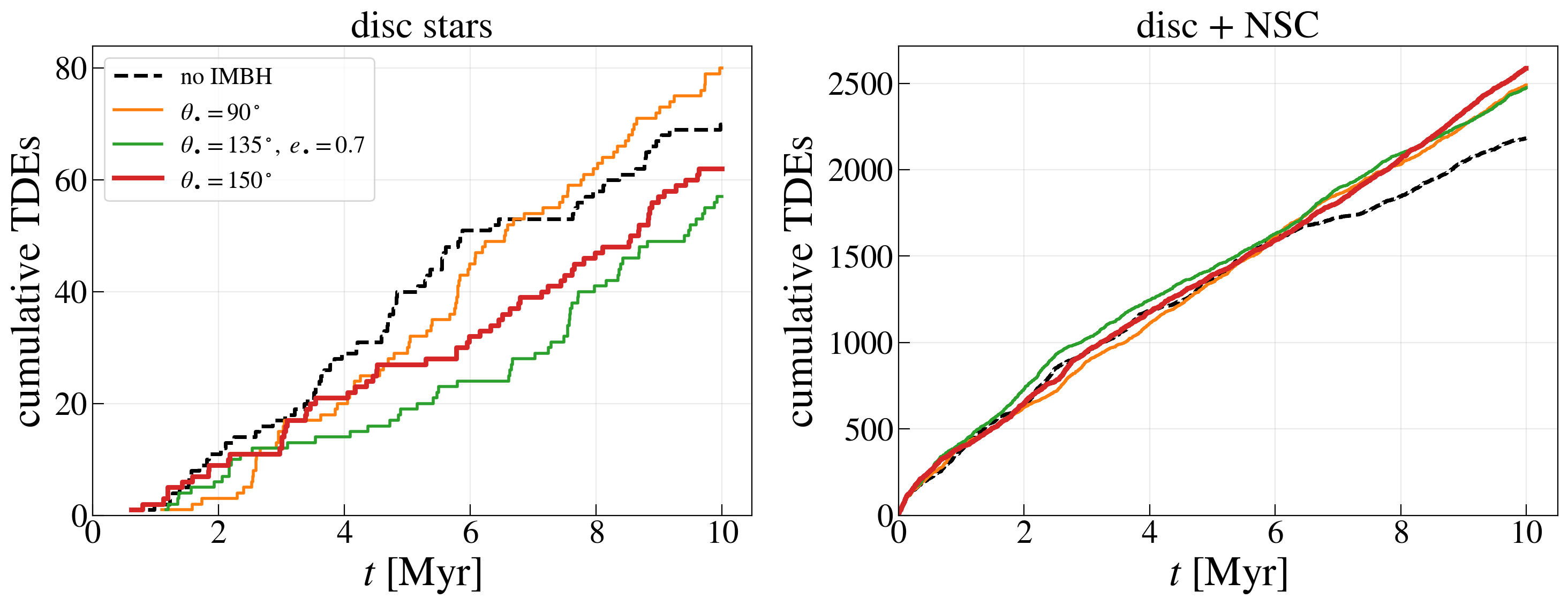}
\caption{Cumulative number of tidal disruptions by the central SMBH as a function of time for the disc stars (left) and for the whole stellar system (right), in the model without an IMBH (black dashed) and in the three $2000\msun$, $a_\IMBH=0.15$~pc IMBH models ($\theta_\IMBH=90^\circ$; $\theta_\IMBH=135^\circ$ with $e_\IMBH=0.7$; and $\theta_\IMBH=150^\circ$; colours as elsewhere in the paper).}
\label{fig:tde_cum}
\end{figure*}

\begin{figure*}
\centering
\includegraphics[width=0.8\linewidth]{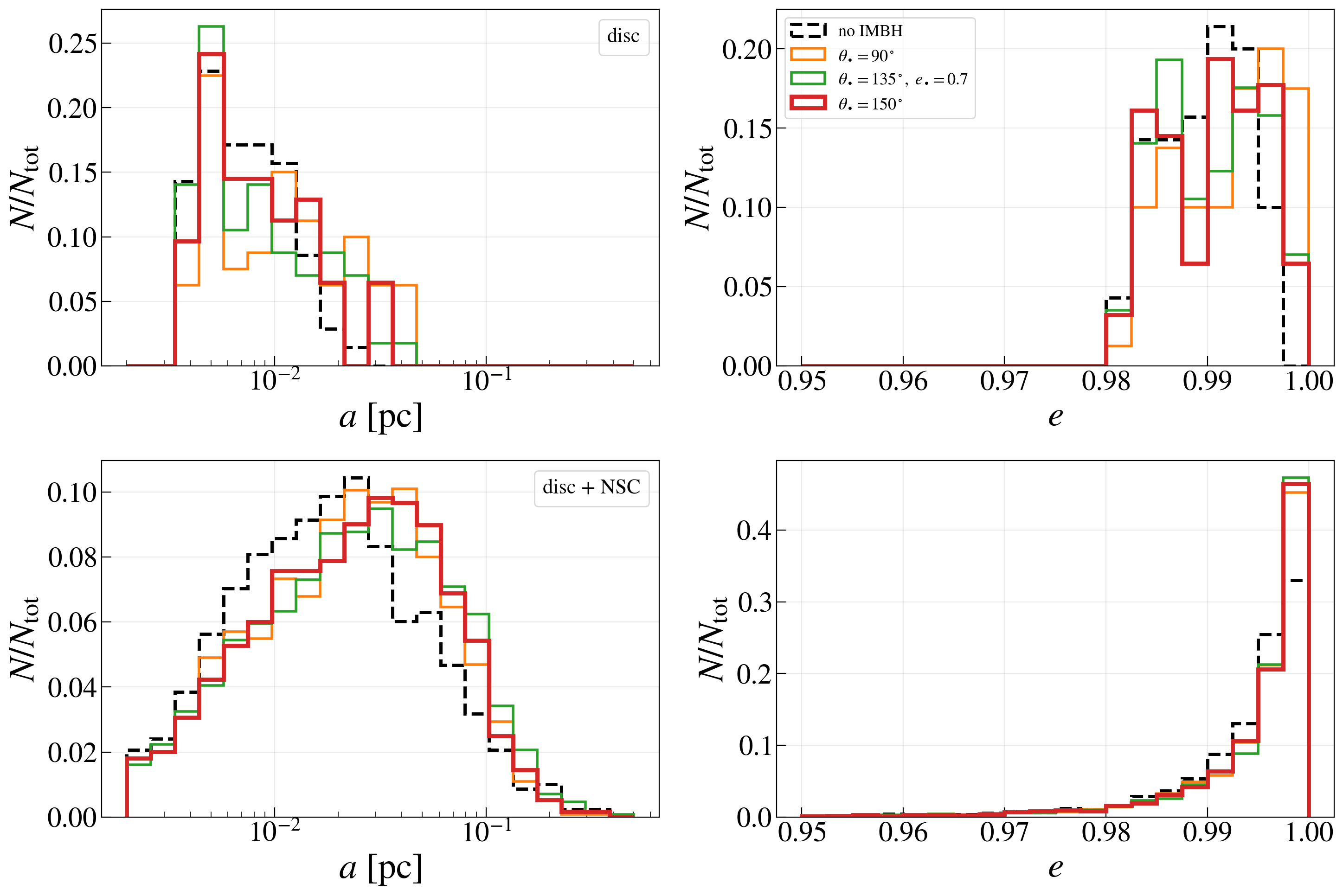}
\caption{Orbital-element distributions of the disrupted stars at the moment of disruption: semi-major axis (left column) and eccentricity (right column), for the disrupted disc stars (top row) and for all disrupted stars (bottom row), in the same models and over the same $0$--$10$~Myr interval as in Fig.~\ref{fig:tde_cum}. Each histogram is normalised to unit sum.}
\label{fig:tde_hist}
\end{figure*}

The central black hole in our simulations grows by consuming stars that pass within the capture radius $r_\mathrm{acc}=6.9\times10^{-5}$~pc ($14.2$~AU; Section~\ref{sec:simulations}). We refer to these captures as tidal disruption events (TDEs). For numerical reasons, we adopt a capture radius much larger than $r_\mathrm{t,\odot}$, the physical tidal radius of a solar-type star: $r_\mathrm{acc}=14.2\,\mathrm{AU}\simeq19\,r_\mathrm{t,\odot}$, where $r_\mathrm{t,\odot}\simeq0.74$~AU. In general, the tidal radius is $r_\mathrm{t}=R_\star(M/m_\star)^{1/3}$ for a star of mass $m_\star$ and radius $R_\star$ \citep{Hills1975, Rees1988}. The same expression gives the disruption radius of a stellar binary when $R_\star$ and $m_\star$ are replaced by the binary separation and mass. The simulated captures therefore allow us to measure how efficiently the disc and the NSC feed the loss cone, defined here as orbits with pericentres inside $r_\mathrm{acc}$, and to estimate stellar tidal disruption rates and the rate of binary separations via the Hills mechanism.

All rates below are averages over $10$~Myr, including the adopted binary fractions for binary separations and Hills captures (Table~\ref{tab:etabeta}).

Figure~\ref{fig:tde_cum} compares cumulative disruptions in the simulation without an IMBH and three simulations with $2000\msun$ and $a_\IMBH=0.15$~pc. The cluster dominates the counts, with $\simeq2200$ disruptions within $10$~Myr without an IMBH and $2500$--$2600$ with one. The $\theta_\IMBH=179^\circ$, $a_\IMBH=0.05$~pc model gives $\simeq3500$.

Although disc stars constitute about 1 per cent of the stellar population, they account for 2--3 per cent of the disruptions. Over 10 Myr, the disrupted fraction of the disc population is 2.4--3.3 times that of the NSC. At $10$~Myr, the polar model has $80$ disrupted disc stars, the two retrograde $a_\IMBH=0.15$~pc models have $57$ and $62$, and the no-IMBH model has $70$, respectively. These differences lie within the Poisson uncertainty.

Disrupted disc stars have semi-major axes $a\simeq0.004$--$0.04$~pc and nearly radial orbits ($e>0.98$; Fig.~\ref{fig:tde_hist}). The IMBH shifts the disrupted population towards larger semi-major axes, most clearly for the NSC stars.

\subsection{Loss-cone regime and dependence on the tidal radius}
\label{subsec:tde-losscone}

\begin{figure*}
\centering
\includegraphics[width=0.8\linewidth]{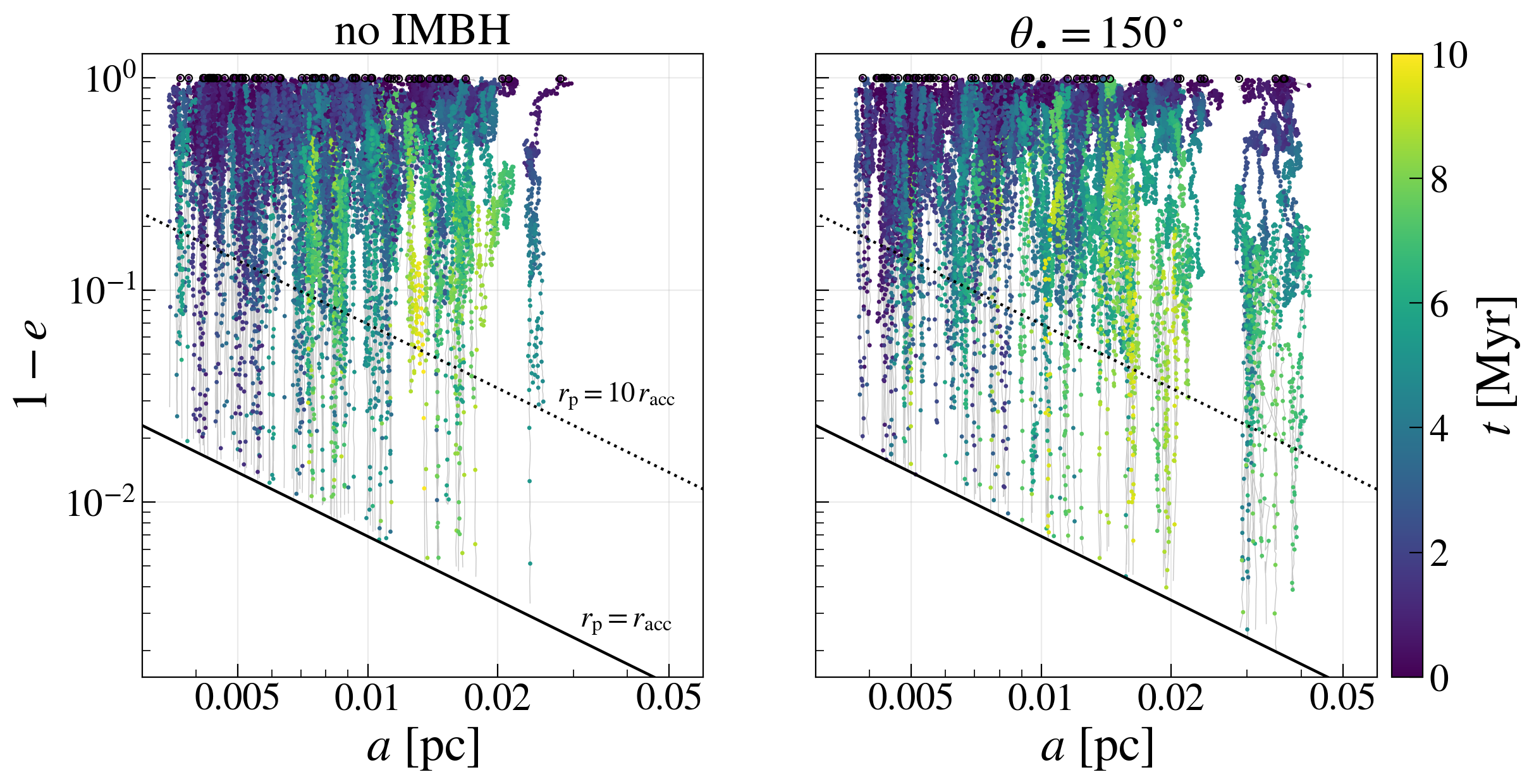}
\caption{Orbits of the disc stars disrupted within $10$~Myr in the model without an IMBH (left, $70$ stars) and in the fiducial model ($\theta_\IMBH=150^\circ$; right, $62$ stars) in the plane of $a$ and $1-e$, from $t=0$ to the last sampled orbit before disruption. Open circles mark the orbits at $t=0$, and the dots are coloured by time. The solid and dotted lines mark orbits with $r_\mathrm{p}=r_\mathrm{acc}$ and $10\,r_\mathrm{acc}$.}
\label{fig:tde_tracks}
\end{figure*}

\subsubsection{Loss-cone regime}
\label{subsec:tde-losscone-regime}

Disrupted disc stars lose angular momentum with little change in orbital energy (Fig.~\ref{fig:tde_tracks}). Their typical fractional change in $a$ is about $8$--$10$ per cent, while $1-e$ falls by two orders of magnitude. This is consistent with the expectation that the variations in energy and angular momentum follow $\delta E/E\sim \delta J/J_c\sim \delta v/v$ where $J_c=(1-e^2)^{-1/2}J=(G M a)^{1/2}$ is the circular angular momentum, implying that the relaxation time in eccentricity or angular momentum $J$ is accelerated by a factor $(1-e^2)^{-1}$ relative to energy or semimajor axis, and coherent torques (scalar resonant relaxation) may give an additional enhancement outside of the Schwarzschild barrier where relativistic precession is subdominant \citep{LightmanShapiro1977,Rauch1996,Hopman2006,Bar-Or2016}.

The loss cone comprises orbits with pericentres inside $r_\mathrm{acc}$. We define its angular-momentum boundary $J_\mathrm{lc}=\sqrt{2GMr_\mathrm{acc}}$ and, following \citet{LightmanShapiro1977}, the loss-cone parameter
\begin{equation}\label{eq:qlc-def}
q_\mathrm{lc}=\frac{\left\langle(\Delta J)^2\right\rangle(P_\mathrm{orb})}{J_\mathrm{lc}^2},
\end{equation}
where $J$ is the magnitude of the specific angular momentum of a star, $\left\langle(\Delta J)^2\right\rangle(\tau)$ the mean squared change of $J$ over a time lag $\tau$, and $P_\mathrm{orb}$ the orbital period. The loss cone is empty for $q_\mathrm{lc}\ll1$ and full for $q_\mathrm{lc}\gg1$ \citep{FR1976, LightmanShapiro1977}.

We compute $\left\langle(\Delta J)^2\right\rangle(\tau)$ by averaging $[J(t+\tau)-J(t)]^2$ over a set of stars and over all start times $t$ within an interval, and fit
\begin{equation}\label{eq:qlc}
\frac{\left\langle(\Delta J)^2\right\rangle}{J_\mathrm{lc}^2}=\eta\,\frac{\tau}{P_\mathrm{orb}}+\left(\beta\,\frac{\tau}{P_\mathrm{orb}}\right)^2,
\end{equation}
where the coefficients $\eta$ and $\beta$ describe diffusion and steady drift per orbit. We minimise squared relative residuals with $\eta,\beta^2\ge0$, giving $q_\mathrm{lc}=\eta+\beta^2$. 
We fit the disc and NSC populations separately in semi-major-axis bins over $10^4$~yr starting at $t=5$~Myr, using the period at each bin centre. This interval corresponds to about $200$ orbital periods at $a=0.01$~pc and one at $a\simeq0.35$~pc. We also fit each disrupted star over the same duration preceding disruption, using the period at its median semi-major axis.

\begin{figure*}
\centering
\includegraphics[width=0.8\linewidth]{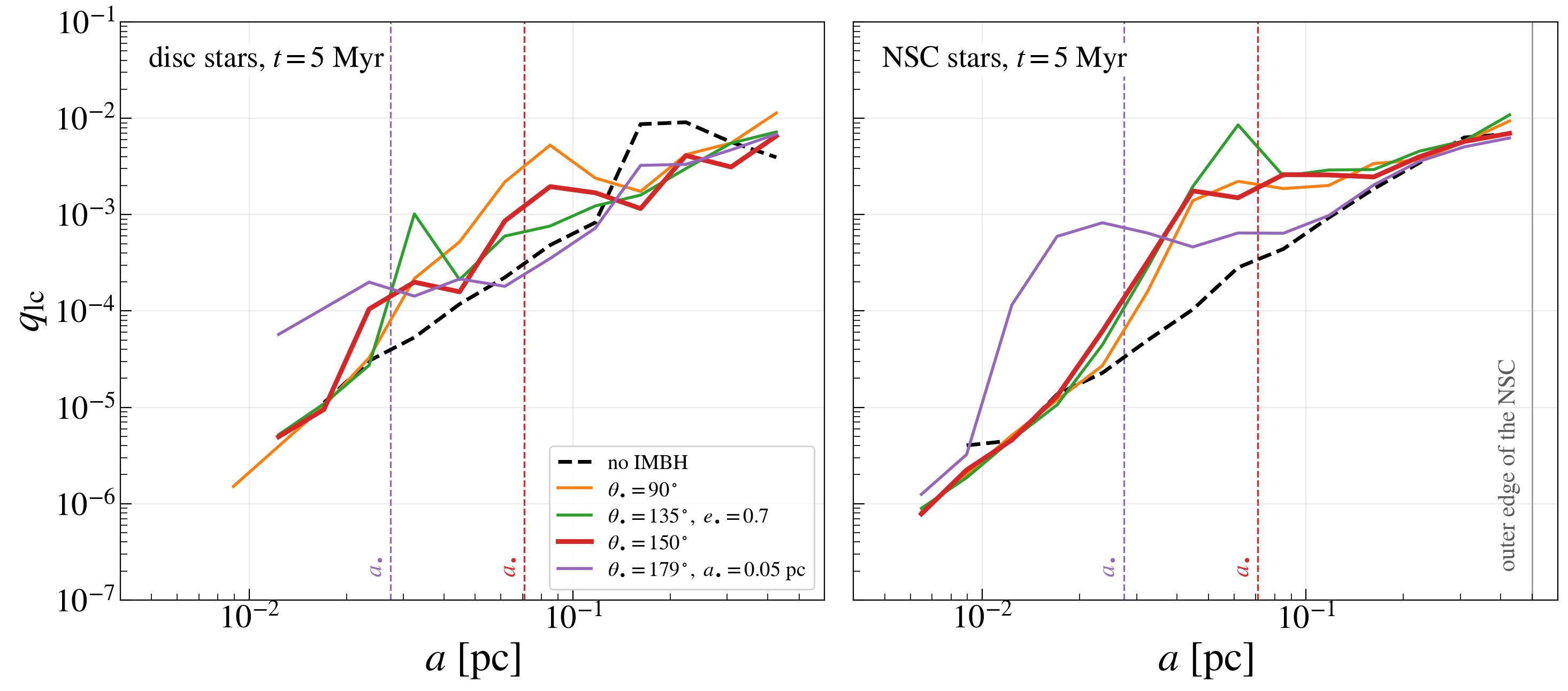}
\caption{Loss-cone parameter $q_\mathrm{lc}$ (Eq.~\eqref{eq:qlc-def}) of the disc stars (left) and of the NSC stars (right) in bins of semi-major axis at $t=5$~Myr. Red and purple dashed vertical lines mark the IMBH semi-major axes at $5$~Myr in the $\theta_\IMBH=150^\circ$ and $179^\circ$ models, respectively. The grey vertical line marks the outer edge of the simulated NSC ($0.5$~pc). The legend shows the models analysed.}
\label{fig:tde_qprofile}
\end{figure*}

\begin{figure}
\centering
\includegraphics[width=\columnwidth]{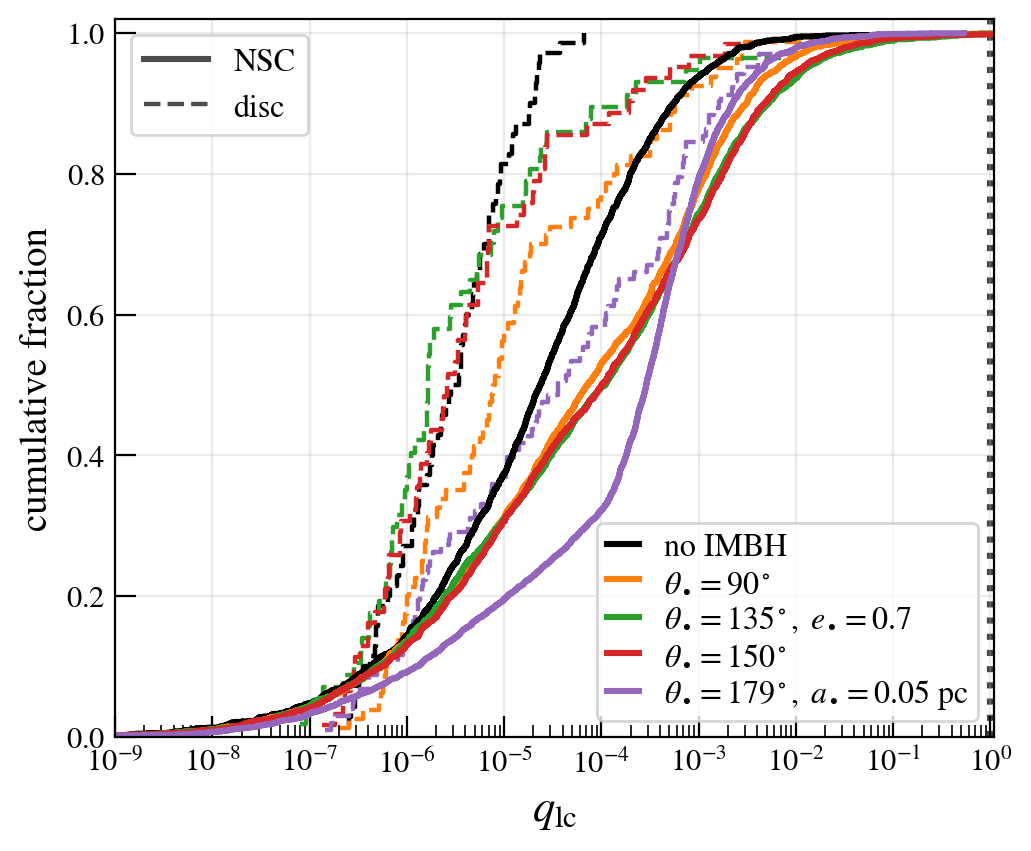}
\caption{Cumulative distributions of $q_\mathrm{lc}$ (Eq.~\eqref{eq:qlc-def}) of the disc stars (dashed) and of the NSC stars (solid) disrupted within $10$~Myr in each model (colours), each measured over the interval preceding its disruption (Section~\ref{subsec:tde-losscone-regime}). NSC stars for which less than two-thirds of this interval is sampled (at most $1.2$ per cent per model) are omitted, and $0.3$ per cent of the disrupted NSC stars lie outside the plotted range. The dotted line marks $q_\mathrm{lc}=1$.}
\label{fig:tde_losscone}
\end{figure}

\begin{table*}
\caption{Mean event rates over $0$--$10$~Myr, followed by the fitted slope $p$ and loss-cone statistics. Binary separations and Hills captures use initially stable binaries with $f_\mathrm{b}=1$ for the disc and $0.025$ for the NSC. Hills captures satisfy the criteria of Section~\ref{subsec:tde-hills}. The slope $p$ describes $N(<R)\propto R^p$ over $r_\mathrm{acc}\le R\le20r_\mathrm{acc}$. The coefficients $\eta$ and $\beta$ are medians for the disrupted stars of Fig.~\ref{fig:tde_losscone}; $f_\beta$ is the fraction with $\beta^2>\eta$ (Eq.~\eqref{eq:qlc}).}
\label{tab:etabeta}
\centering
\small
\setlength{\tabcolsep}{4pt}
\begin{tabular}{llrrrrrrr}
\hline
 & & \multicolumn{3}{c}{Mean rate [Myr$^{-1}$]} & & & & \\
\cline{3-5}
$\theta_\IMBH$ [$^\circ$] & Population & Solar-type TDEs & Binary separations & Hills captures & $p$ & $\eta$ [$10^{-6}$] & $\beta$ [$10^{-3}$] & $f_\beta$ \\
\hline
no IMBH & disc & 5.12 & 8.81 & 2.32 & 0.11 & 0.72 & 1.4 & 0.70 \\
 & NSC & 129 & 6.68 & 1.82 & 0.17 & 8.1 & 3.3 & 0.62 \\
\noalign{\smallskip}
90 & disc & 5.48 & 11.4 & 3.1 & 0.13 & 3.7 & 1.9 & 0.50 \\
 & NSC & 142 & 8.05 & 2.17 & 0.18 & 37 & 4.6 & 0.41 \\
\noalign{\smallskip}
135 & disc & 3.55 & 8.54 & 1.98 & 0.16 & 0.85 & 0.72 & 0.30 \\
 & NSC & 138 & 8.89 & 2.2 & 0.19 & 48 & 5.5 & 0.41 \\
\noalign{\smallskip}
150 & disc & 3.64 & 9.05 & 2.19 & 0.18 & 1.7 & 0.99 & 0.34 \\
 & NSC & 151 & 8.75 & 2.31 & 0.17 & 49 & 5.5 & 0.45 \\
\noalign{\smallskip}
179 & disc & 8.38 & 11.3 & 4.02 & 0.07 & 11 & 3.9 & 0.54 \\
 & NSC & 232 & 9.04 & 3.04 & 0.12 & 87 & 11 & 0.58 \\
\hline
\end{tabular}
\par\medskip
\begin{flushleft}\textbf{Notes.} The IMBH models have $m_\IMBH=2000\msun$, $a_\IMBH=0.15$~pc and $e_\IMBH=0.33$, except $e_\IMBH=0.7$ for $\theta_\IMBH=135^\circ$ and $a_\IMBH=0.05$~pc for $\theta_\IMBH=179^\circ$.\end{flushleft}
\end{table*}

The population fits give $q_\mathrm{lc}<0.012$ at $5$~Myr (Fig.~\ref{fig:tde_qprofile}, and see Table~\ref{tab:etabeta} for $\eta$ and $\beta$). Thus the simulated system is in the empty-loss-cone regime. All disrupted disc stars and about $99\%$ of the disrupted NSC stars in the simulation have $q_\mathrm{lc}<0.1$ (Fig.~\ref{fig:tde_losscone}).

Incoherent diffusion dominates the population fits at $5$~Myr, while coherent drift dominates for $30\%$--$70\%$ of the individual disrupted stars (Table~\ref{tab:etabeta}). The IMBH increases the median coherent NSC drift coefficient. 

\subsubsection{Dependence on the tidal radius}
\label{subsec:tde-radius}

\begin{figure*}
\centering
\includegraphics[width=0.8\linewidth]{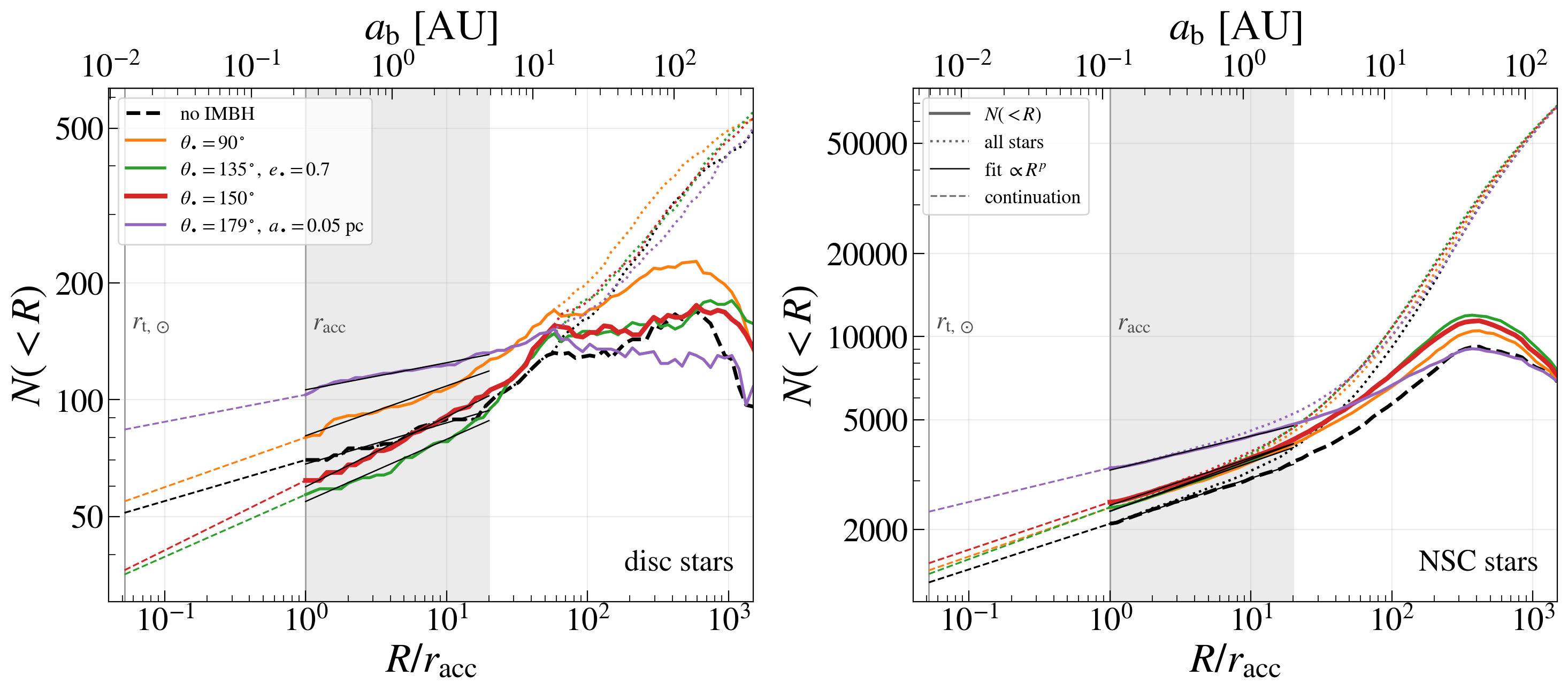}
\caption{Number of stars whose pericentre falls below a radius $R$ within $10$~Myr, for the disc stars (left) and the NSC stars (right). Solid and dashed lines show $N(<R)$, which counts only stars with $r_\mathrm{p}>R$ at $t=0$, and dotted lines of the same colours all stars. Thin black lines show the fits $N\propto R^{p}$ over the shaded range $r_\mathrm{acc}\le R\le20\,r_\mathrm{acc}$, and thin dashed lines in the model colours the extrapolation anchored at $N(<r_\mathrm{acc})$ to the tidal radius of a solar-type star, $r_\mathrm{t,\odot}$. The upper axes give the semi-major axis $a_\mathrm{b}$ of a binary of mass $m_\mathrm{b}=20\msun$ (left) or $2\msun$ (right) with $r_\mathrm{t,b}=R$ (Section~\ref{subsec:tde-binaries}). The legend of the left panel gives the models and that of the right panel the line styles.}
\label{fig:tde_radius}
\end{figure*}

We count first crossings of each pericentre radius $R\ge r_\mathrm{acc}$ within $10$~Myr. We count separately $N(<R)$ including or excluding stars with $r_{\rm p}(0)\le R$, i.e. initial pericentres at the start of the simulation inside $R$, as shown by dotted lines or curves in the model styles, respectively, in Figure~\ref{fig:tde_radius}. Fitting $N\propto R^p$ in logarithmic space over $r_\mathrm{acc}\le R\le20r_\mathrm{acc}$ gives $p=0.07$--$0.18$ for the disc and $0.12$--$0.19$ for the NSC, respectively (Table~\ref{tab:etabeta}).

Assuming the measured slope continues below $r_\mathrm{acc}$, we extrapolate to solar-type stars using $N_\odot=N(<r_\mathrm{acc})(r_\mathrm{t,\odot}/r_\mathrm{acc})^p$. The combined disc and NSC counts within $10$~Myr are $1340$ without an IMBH, $1420$--$1550$ for the $a_\IMBH=0.15$~pc models, and $2400$ for $\theta_\IMBH=179^\circ$. These correspond to rates of $130$--$240$~Myr$^{-1}$.

\subsection{Binary separations}
\label{subsec:tde-binaries}

\begin{figure*}
\centering
\includegraphics[width=0.8\linewidth]{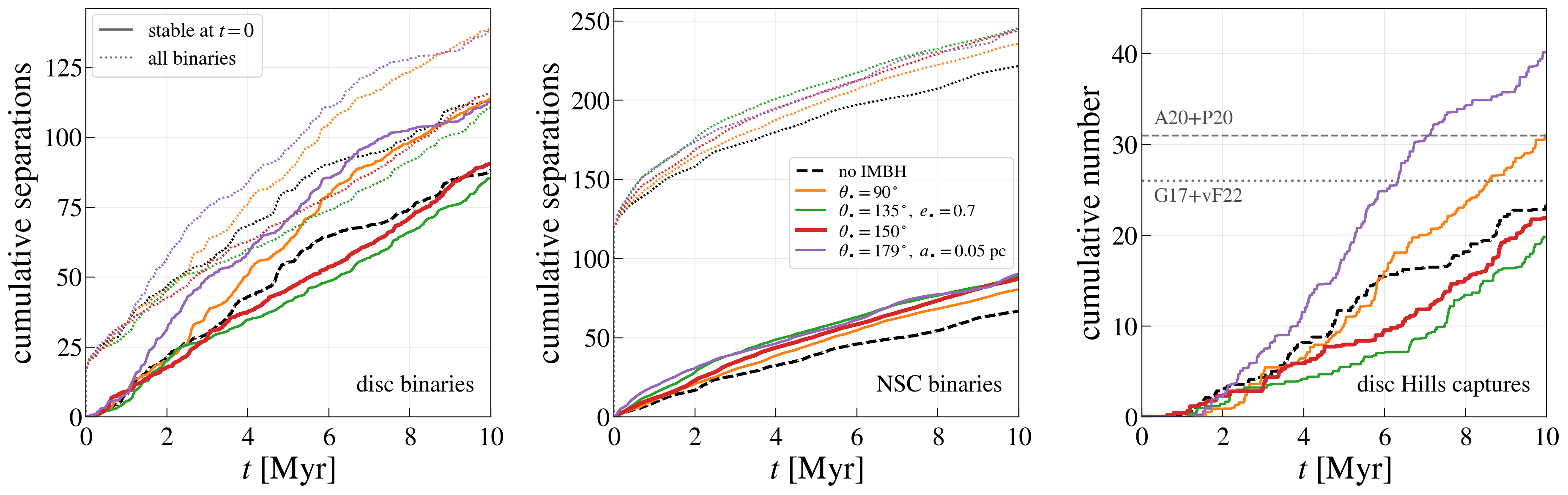}
\caption{Cumulative numbers of separated binaries among the disc stars (left) and the NSC stars (middle), and of disc Hills captures in the S-star region (right; Section~\ref{subsec:tde-hills}), for binary fractions $f_\mathrm{b}=1$ in the disc and $0.025$ in the NSC. Lines in the model styles count the binaries stable at $t=0$, and dotted lines all binaries; the Hills captures are the same for both. The horizontal lines in the right panel mark the observed stars with orbital solutions and $a\le0.04$~pc in the G17$+$vF22 ($26$, dotted) and A20$+$P20 ($31$, dashed) samples (Section~\ref{subsec:sstar-obs}). The model legend applies to all panels.}
\label{fig:tde_binaries}
\end{figure*}

A binary of total mass $m_\mathrm{b}$ and relative semi-major axis $a_\mathrm{b}$ separates when its centre-of-mass pericentre falls below
\begin{equation}
r_\mathrm{t,b}=a_\mathrm{b}\left(\frac{M}{m_\mathrm{b}}\right)^{1/3}
\end{equation}
\citep{Hills1988, YuTremaine2003, Sari2010}. We assign each simulated star a binary with $m_\mathrm{b}=20\msun$ (disc) or $2\msun$ (NSC). We adopt a distribution uniform in $\ln a_\mathrm{b}$ \citep{Opik1924, Duchene2013}, from contact ($0.05$ and $0.01$~AU, respectively) to the Hill radius $a_\mathrm{H}=r(m_\mathrm{b}/3M)^{1/3}$ at distance $r$ from the SMBH \citep{Chu2018, Rose2020}. We evaluate this upper limit at $r=0.04$~pc for the disc and $0.1$~pc for the NSC, the semi-major axes enclosing most separations.

We average the crossing counts over this distribution and multiply by the binary fraction $f_\mathrm{b}$. We adopt $f_\mathrm{b}=1$ for the disc and $f_\mathrm{b}=0.025$ for the NSC \citep{Panamarev2019}. Below $r_\mathrm{acc}$, we use the extrapolation of Section~\ref{subsec:tde-radius}. 

Their mean separation rates are $8.5$--$11.4$~Myr$^{-1}$ in the disc and $6.7$--$9.0$~Myr$^{-1}$ in the NSC (Table~\ref{tab:etabeta}). Including initially unstable binaries gives upper limits of $11$--$14$ and $22$--$25$~Myr$^{-1}$, respectively. Binaries initially inside $r_\mathrm{t,b}$ separate on their first pericentre passage. Figure~\ref{fig:tde_binaries} shows the cumulative counts as a function of time.

\subsection{Hills captures in the S-star region}
\label{subsec:tde-hills}

For an equal-mass circular binary, we estimate the specific energy change as $\Delta E=v_\mathrm{cm}\delta v$, with $v_\mathrm{cm}\simeq(2GM/r_\mathrm{t,b})^{1/2}$ and $\delta v=(Gm_\mathrm{b}/a_\mathrm{b})^{1/2}/2$ \citep{YuTremaine2003}. This assumes a nearly parabolic encounter, with the velocities relative to the binary centre of mass parallel and antiparallel to its motion about the SMBH. Retaining the initial binding energy, the more tightly bound star has a very high eccentricity as
\begin{equation}\label{eq:hills}
\begin{aligned}
\frac{1}{a_\mathrm{cap}}&=\frac{1}{a}+\frac{1}{a_\mathrm{cap,0}}, \qquad
a_\mathrm{cap,0}=\frac{a_\mathrm{b}}{\sqrt{2}}\left(\frac{M}{m_\mathrm{b}}\right)^{2/3}, \\
e_\mathrm{cap}&=1-\frac{r_\mathrm{t,b}}{a_\mathrm{cap}} \\
&=1-\frac{r_\mathrm{t,b}}{a}-\sqrt{2}\left(\frac{m_\mathrm{b}}{M}\right)^{1/3}.
\end{aligned}
\end{equation}
where $a$ is the centre-of-mass semimajor axis immediately before separation, and $a_\mathrm{cap,0}=GM/(2\Delta E)$ is the capture semi-major axis for a parabolic encounter. We approximate the captured star's pericentre by $r_\mathrm{t,b}$ \citep{Kobayashi2012, Generozov2020}. We count a Hills capture in the S-star region when the companion escapes ($a_\mathrm{cap,0}<a$) and $a_\mathrm{cap}\le0.04$~pc. Table~\ref{tab:etabeta} gives the mean capture rates for both populations; the NSC entries refer to solar-mass stars.

For the disc binaries with $f_\mathrm{b}=1$, we predict $6$--$17$ such captures within $5$~Myr and $20$--$40$ within $10$~Myr (Fig.~\ref{fig:tde_binaries}). Their initial eccentricities exceed $0.95$, with a median of $0.97$. The $10$-Myr count is comparable to the $20$ confirmed early-type S-stars with $a\le0.04$~pc in the G17$+$vF22 sample (Section~\ref{subsec:sstar-obs}). 

The predicted counts scale with $f_\mathrm{b}$ and depend on the adopted energy change and binary distribution. Adopting the smaller energy changes considered by \citet{Generozov2020} reduces our $10$-Myr counts by factors of $1.2$--$2.9$. Truncating the binary distribution at $10$~AU instead of $a_\mathrm{H}$ increases them by a factor of $1.43$.

In each model, $91\%$--$100\%$ of captures come from binaries born inside $0.04$~pc, mainly redistributing stars already in the S-star region. For binaries born at $0.04$--$0.5$~pc, the $10$-Myr capture efficiencies are below $0.4\%$ on disc orbits and $0.4$--$0.7$ per cent on isotropic NSC orbits. These captures can therefore supply highly eccentric stars to the S-star region.

%###########################################################################
\section{Summary and discussion}
\label{sec:SUM}
%###########################################################################

We have presented direct $N$-body simulations of a young stellar disc around the central SMBH, perturbed by an IMBH and embedded in a live NSC represented by $10^{5}$ particles. In \PaperI, we showed that an inclined IMBH can break an initially coherent stellar disc into multiple structures in angular-momentum space. In that study, the surrounding spherical NSC was represented by a smooth static potential, which suppressed both the stochastic torques associated with two-body and resonant relaxation and dynamical friction on the IMBH. Consequently, the evolutionary timescales obtained in \PaperI were expected to be upper limits. Here we included these processes self-consistently and determined whether the resulting structures  develop within the observed age of the Galactic-centre stellar disc.

Our main results are as follows.

\begin{itemize}

\item \emph{An inclined IMBH produces disc substructure similar to that observed within the age of the young stellar population.}
In our fiducial retrograde model ($\theta_\IMBH=150^\circ$, $a_\IMBH=0.15$~pc, $e_\IMBH=0.33$, $m_\IMBH=2000\msun\simeq0.67\,M_\mathrm{d}$), the IMBH breaks the stellar disc into three disc-like components and a filament in angular-momentum space by $t\simeq5$~Myr in a configuration similar to the observed Galactic-centre substructures. In the static background of \PaperI, comparable separations developed only after $10$--$20$~Myr. The stochastic torques of the live NSC, together with the inward migration of the IMBH, accelerate the angular separation between the substructures by a factor of a few, bringing it within the estimated age of the young stars ($6\pm2$~Myr, \citealt{PaumardEtAl2006}; $2.5$--$5.8$~Myr, \citealt{Lu2013}). The two inner components, which are misaligned by $131^\circ$, similarly to the clockwise disc and the counter-clockwise structure (${\simeq}130^\circ$; \citealt{vonFellenberg2022}), have median eccentricities ($e\simeq0.42$ and $0.36$) comparable to those inferred for the observed disc components.

\item \emph{The innermost disc stars evolve into an S-star-like population, largely through relaxation in the live NSC.}
Stars with $a\le0.04$~pc develop broad distributions of inclination, node and eccentricity that overlap those of the observed S-stars within $5$--$10$~Myr. Much of this evolution occurs even without an IMBH. In the no-IMBH model the inner disc thickens within $10$~Myr towards a nearly spherical distribution, with $\Delta\iota\simeq40^\circ$--$50^\circ$ (an isotropic distribution has $\Delta\iota=54.7^\circ$; Section~\ref{subsec:warp}), the orientations of the innermost stars disperse, and their eccentricities grow.
The fraction of innermost stars inside the `zone of avoidance' \citep{Burkert2024} falls from $100$\% to the observed $8$--$23$\% range within $\sim10$~Myr in every simulation, including the one without an IMBH.

\item \emph{The live cluster also changes the orbit of the IMBH.}
Dynamical friction causes the massive IMBHs
($m_\IMBH=2000\msun$) initially at $a_\IMBH=0.15$~pc to sink by more than a factor of two within $10$~Myr. The measured decay is reproduced by the Chandrasekhar prediction of Eq.~\eqref{eq:df}, evaluated using the measured cluster profile, when the standard Coulomb logarithm
$\ln\Lambda=\ln(M/m_\IMBH)$ is reduced by a factor of two. As in \PaperI, prograde IMBHs align with the radially overlapping disc stars, whereas massive retrograde IMBHs tend to anti-align. The initially orthogonal configuration is no longer an equilibrium in the live cluster, and the $\theta_\IMBH=90^\circ$ model aligns with the overlapping stars within $\sim8$~Myr.

\item \emph{Tidal disruption rates are dominated by the cluster and modulated by the IMBH.}
At the simulated capture radius, about $2200$ stars are disrupted within $10$~Myr in the no-IMBH model, $2500$--$2600$ in the $a_\IMBH=0.15$~pc polar and retrograde models and $3500$ in the $a_\IMBH=0.05$~pc retrograde model. These counts correspond to enhancements of $14$--$19\%$ and $60\%$, respectively. The stellar disc contributes only a few per cent of the events. Extrapolation to the solar tidal radius gives total TDE rates of $130$--$240$~Myr$^{-1}$, averaged over $0$--$10$~Myr. For initially stable binaries, the mean separation rates are $8.5$--$11.4$~Myr$^{-1}$ in the disc and $6.7$--$9.0$~Myr$^{-1}$ in the NSC (Table~\ref{tab:etabeta}). These estimates use binary fractions of $1$ and $0.025$, respectively. The corresponding Hills-capture rates inside $0.04$~pc are $2.0$--$4.0$~Myr$^{-1}$ for the massive disc binaries and $1.8$--$3.0$~Myr$^{-1}$ for the solar-mass components of NSC binaries. Almost all disc Hills captures originate from binaries born inside $0.04$~pc and initially have $e\gtrsim0.95$.
\end{itemize}

The central result of our live-cluster calculations is therefore twofold. First, they confirm the basic mechanism driving the angular momentum distribution identified in \PaperI while bringing its timescale into agreement with the age of the Galactic-centre young stars. A single inclined IMBH with $m_\IMBH\simeq0.67\,M_\mathrm{d}$ initially at $a_\IMBH\simeq0.15$~pc can produce the strongly misaligned, multi-component angular-momentum structure within $\sim5$~Myr. Second, they separate the signatures of the IMBH from those generated by the NSC itself. Relaxation against the cluster alone produces substantial evolution of the innermost stars, including thickening of the inner disc, dispersal of their orbital planes, growth of their eccentricities and depletion of the zone of avoidance, while the IMBH further accelerates the orbital-plane reorientation. The fragmentation of the main body of the disc into well-separated components, together with the additional eccentricity excitation in the radial range traversed by the IMBH, provides a cleaner diagnostic of an intermediate-mass companion.

\citet{Zheng2026} attribute the kinematics of the young stars to a more massive (${\sim}10^{4}\msun$) companion outside the disc ($a\sim0.35$~pc), aided by a sweeping secular resonance in a depleting gaseous disc. Our model and that of \citet{Zheng2026} both invoke an intermediate-mass companion and produce broadly comparable, moderate eccentricities in the clockwise-disc region ($e\sim0.3$--$0.4$ at $a\sim0.1$~pc). The main difference between the two studies is the evolution of the disc substructures. In our simulations, the outer substructures are coherent and remain intact, while the inner ones remain prominent for up to $6$~Myr but are subsequently dispersed by relaxation. In \citet{Zheng2026}, by contrast, the counter-rotating component is transient, and its stars only episodically cluster into a disc-like structure. 
Establishing whether the observed off-disc stars constitute a long-lived coherent structure or a transient configuration may therefore help distinguish between the two scenarios.

Our simulations follow orbital dynamics with stellar masses evolved to $1$~Gyr and further mass-loss evolution is neglected \citep{Panamarev2022}. The comparisons use all selected disc particles, a population that can include black holes, other remnants and faint stars. The resulting distributions describe the dynamics of this population, while observed young-star samples also depend on stellar luminosities, lifetimes and selection effects.

Following stellar evolution and binaries directly with codes such as \textsc{nbody6++gpu} \citep{Wang2015,Spurzem2023} would make our predictions for observable substructure and S-star formation more realistic. The highly eccentric stars supplied by binary separations could then be followed through their subsequent relaxation and stellar evolution. Such calculations could track how stellar mass loss modifies the substructures and which stars remain observable as their luminosities evolve. Future work could also include the tidal-disruption prescription of \citet{Cho2026}, which accounts for the orbit-dependent fraction of stellar debris accreted by the SMBH. The evolution of the angular-momentum components should also be examined in rotating and aspherical NSCs, where additional torques may compete with those from the IMBH. In particular, cluster rotation would allow a direct comparison with the resonant-friction mechanism of \citet{Levin2024}. Torques from the circumnuclear gas disc could further reorient the outer component \citep{Subr+2009,Haas_Subr_Kroupa2011,Kocsis2011}.

%###########################################################################
\section{Acknowledgements}
%###########################################################################

We acknowledge the support of the Project No. BR34836926 ``Research and monitoring of near-Earth and deep space through the development and scaling of the Kazakhstan optical telescope network'', financed by the Aerospace committee of the Ministry of Artificial Intelligence and Digital Development of the Republic of Kazakhstan. This work was supported by the Science and Technology Facilities Council Grant Number ST/W000903/1.

\section*{Data Availability}
The data underlying this article will be shared on reasonable request to the corresponding author.

\bibliography{thesis}

%###########################################################################
\appendix
%###########################################################################

\section{Sky-map evolution of the disc}
\label{app:skyev}

\begin{figure*}
    \centering
    \includegraphics[width=0.85\linewidth]{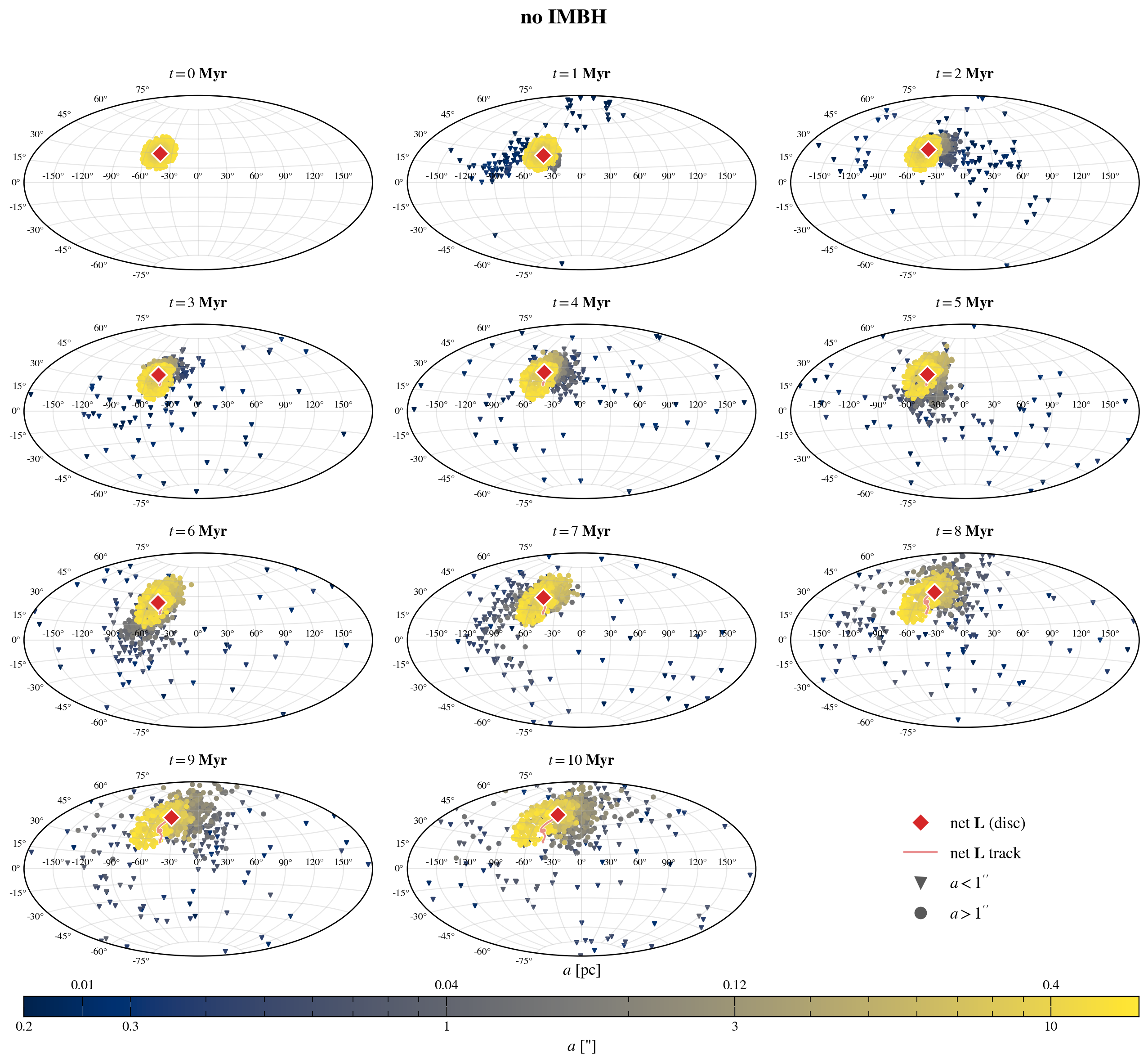}
    \caption{Directions of the disc stars' orbital angular-momentum vectors on
    an Aitoff projection for the model without an IMBH, from $t=0$ to $10$~Myr
    in $1$~Myr steps. Colours give the semi-major axis (colour bar in
    arcseconds, with the corresponding parsec scale above); downward triangles
    mark the innermost stars ($a\le0.04$~pc, $1''$). The red diamond is the net
    angular-momentum direction of the disc, and the faint red track the path it
    follows from $t=0$ to the epoch shown; in the models with an IMBH the
    diamond gives the net direction of the disc and the IMBH together, the
    black star marks the IMBH, and a second faint track follows it over the
    same interval. The remaining panel of each grid repeats this key.
    Projection, rotation and colour scale are as in Fig.~\ref{fig:skymaps}.}
    \label{fig:skyev_noimbh}
\end{figure*}

\begin{figure*}
    \centering
    \includegraphics[width=0.85\linewidth]{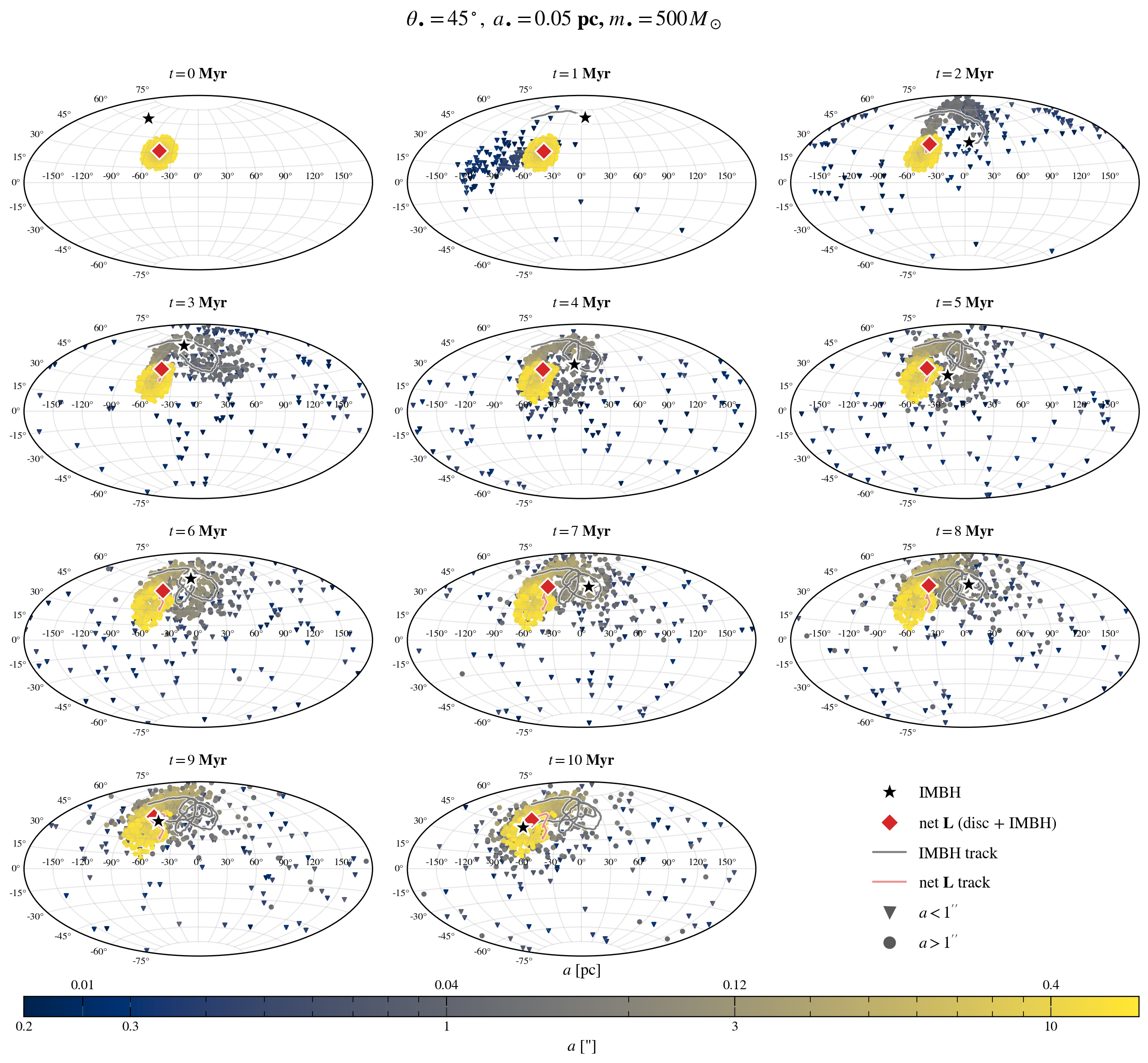}
    \caption{As Fig.~\ref{fig:skyev_noimbh}, for the prograde
    $\theta_\IMBH=45^\circ$, $a_\IMBH=0.05$~pc, $m_\IMBH=500\msun$ model.}
    \label{fig:skyev_i45m500}
\end{figure*}

\begin{figure*}
    \centering
    \includegraphics[width=0.85\linewidth]{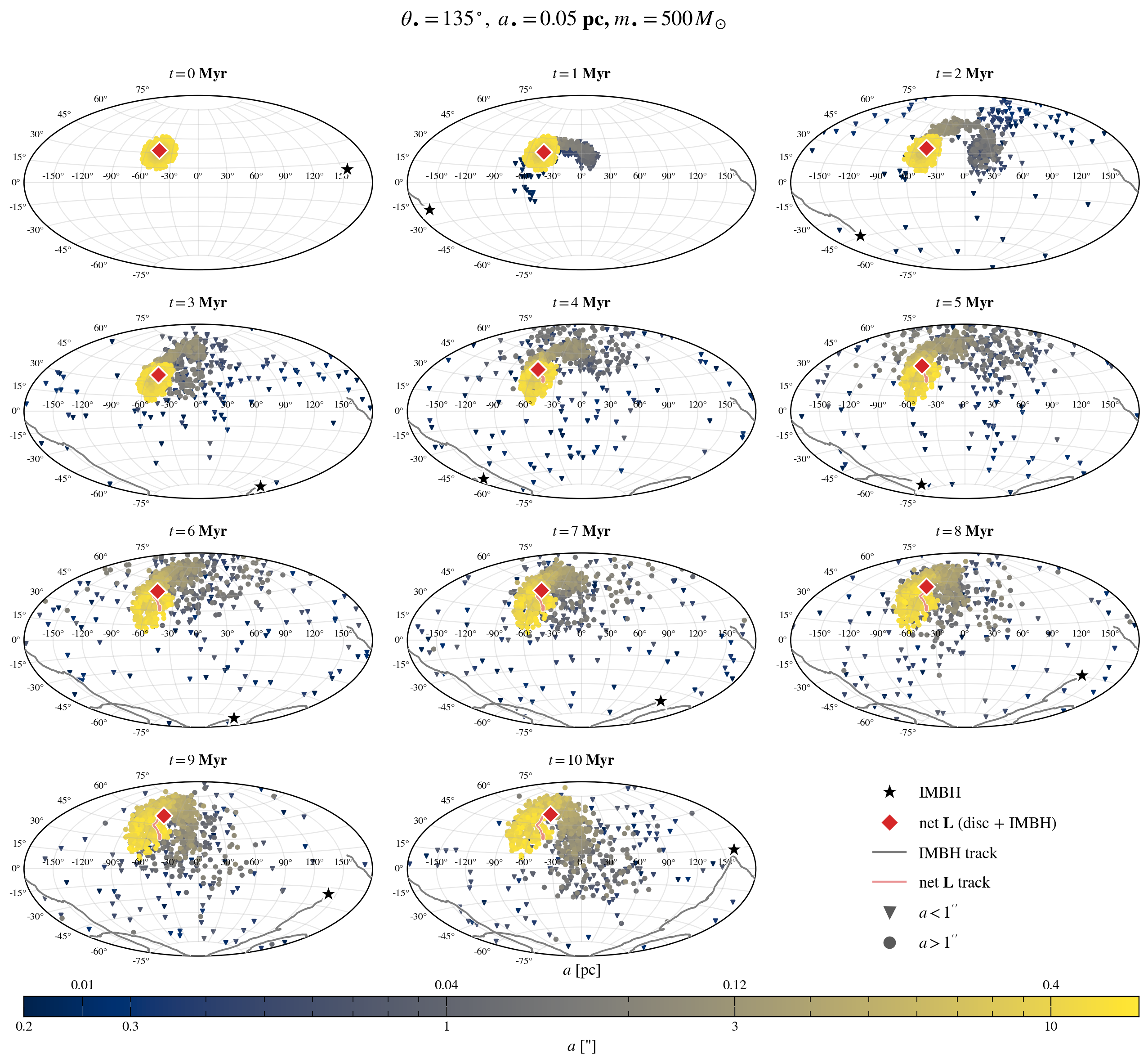}
    \caption{As Fig.~\ref{fig:skyev_noimbh}, for the retrograde
    $\theta_\IMBH=135^\circ$, $a_\IMBH=0.05$~pc, $m_\IMBH=500\msun$ model.}
    \label{fig:skyev_i135m500}
\end{figure*}

\begin{figure*}
    \centering
    \includegraphics[width=0.85\linewidth]{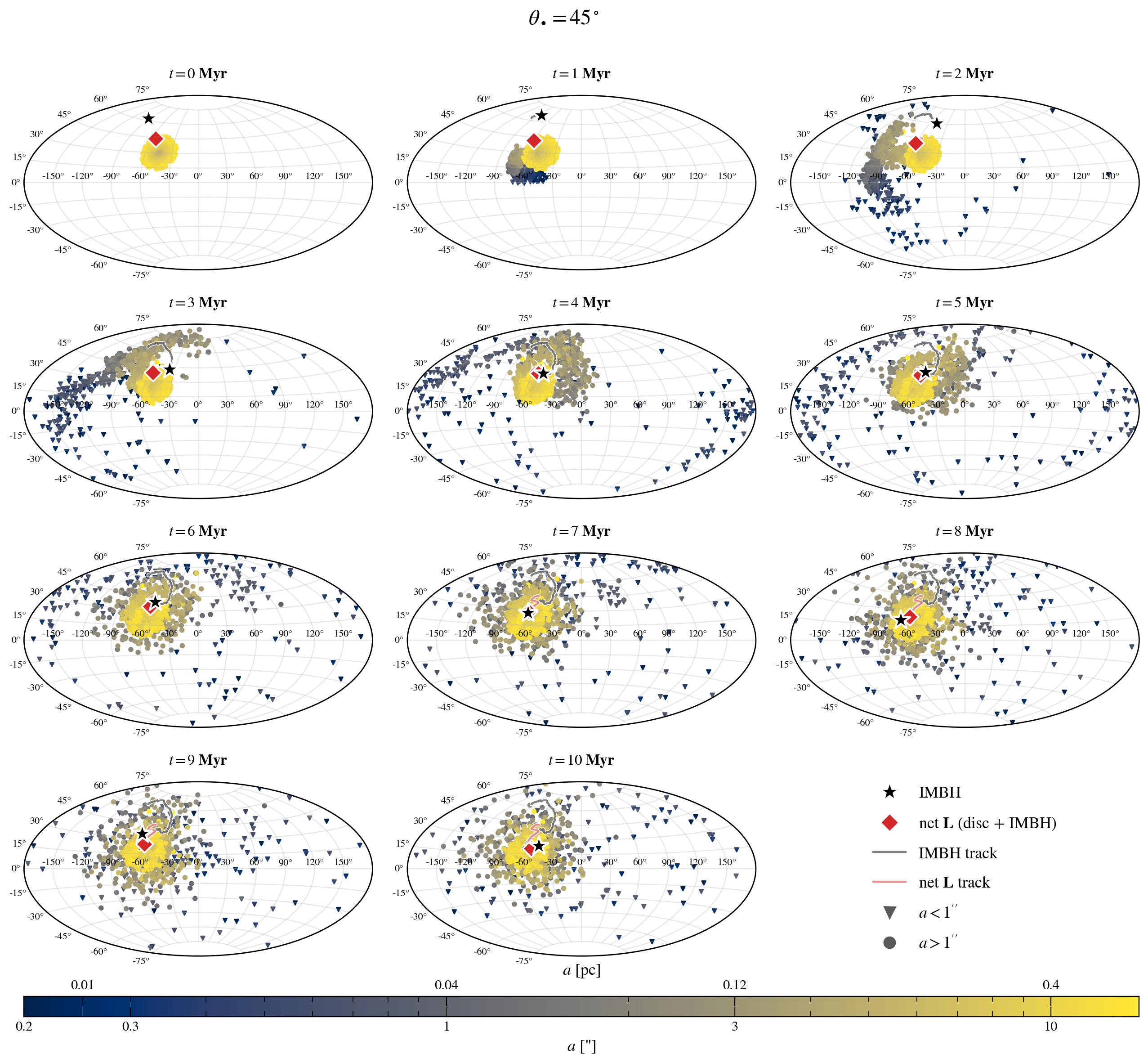}
    \caption{As Fig.~\ref{fig:skyev_noimbh}, for the prograde
    $\theta_\IMBH=45^\circ$, $a_\IMBH=0.15$~pc, $m_\IMBH=2000\msun$ model.}
    \label{fig:skyev_i45}
\end{figure*}

\begin{figure*}
    \centering
    \includegraphics[width=0.85\linewidth]{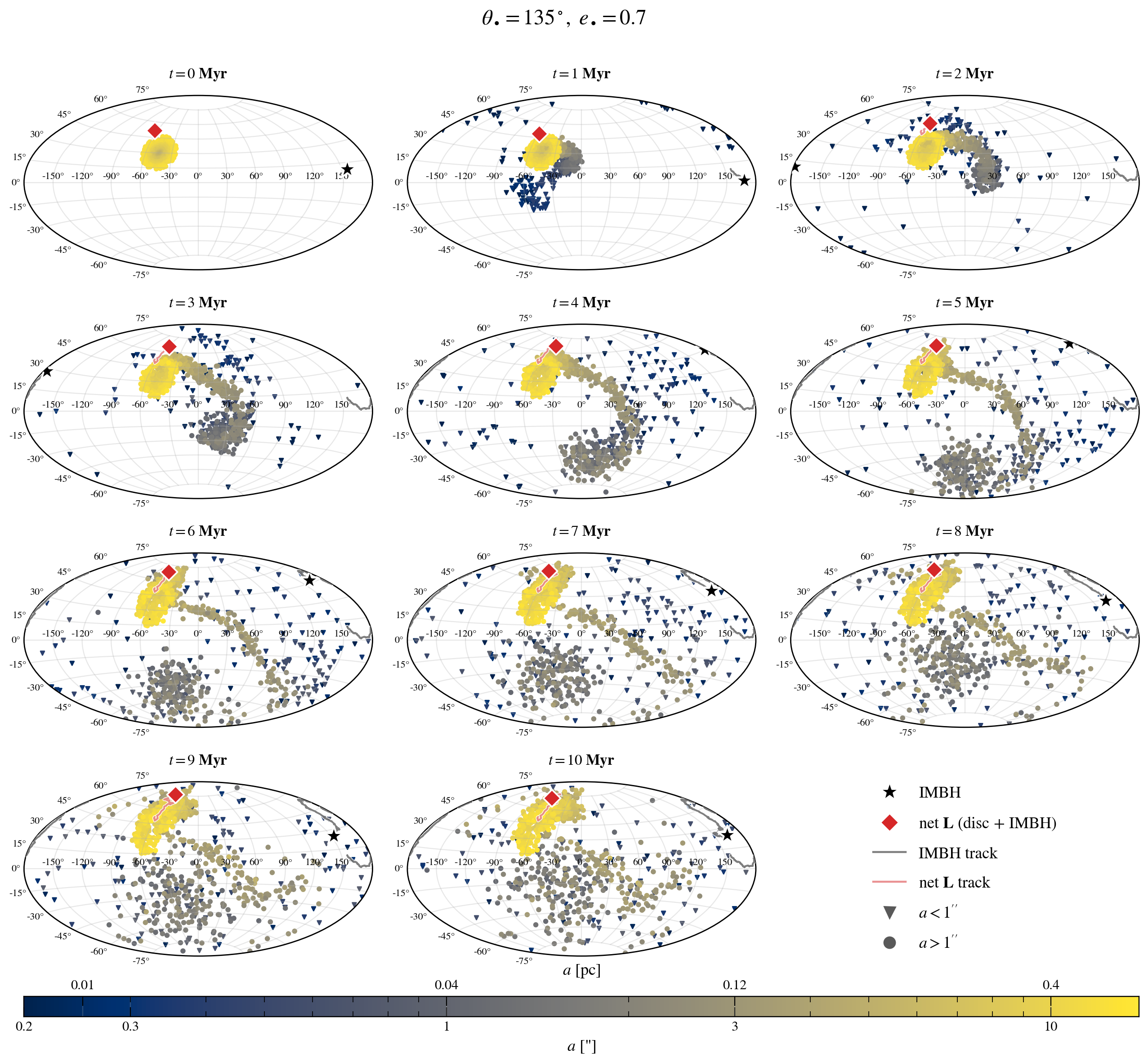}
    \caption{As Fig.~\ref{fig:skyev_noimbh}, for the retrograde
    $\theta_\IMBH=135^\circ$, $a_\IMBH=0.15$~pc, $e_\IMBH=0.7$,
    $m_\IMBH=2000\msun$ model.}
    \label{fig:skyev_i135}
\end{figure*}

\begin{figure*}
    \centering
    \includegraphics[width=0.85\linewidth]{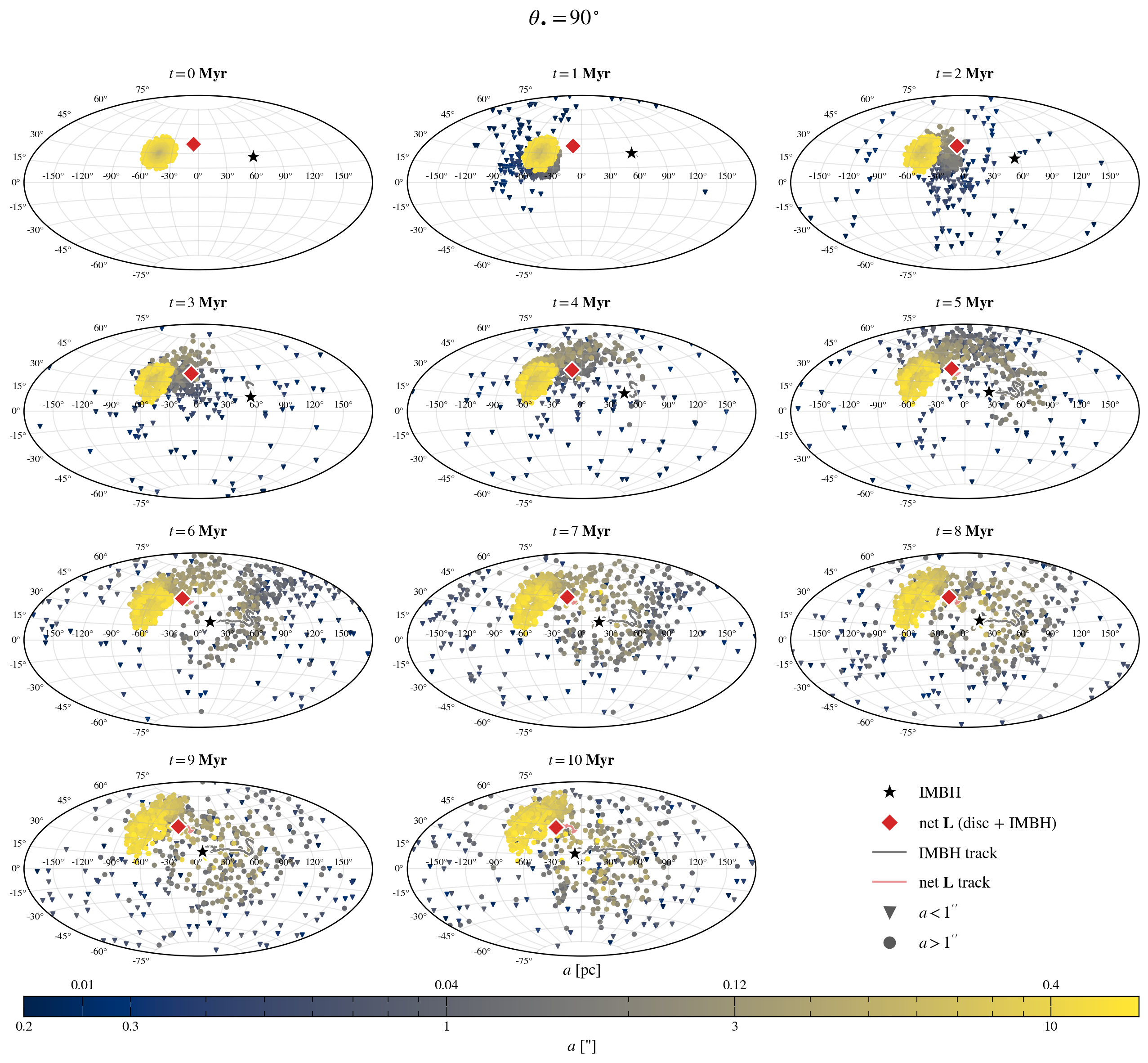}
    \caption{As Fig.~\ref{fig:skyev_noimbh}, for the polar
    $\theta_\IMBH=90^\circ$, $a_\IMBH=0.15$~pc, $m_\IMBH=2000\msun$ model.}
    \label{fig:skyev_i90}
\end{figure*}

\begin{figure*}
    \centering
    \includegraphics[width=0.85\linewidth]{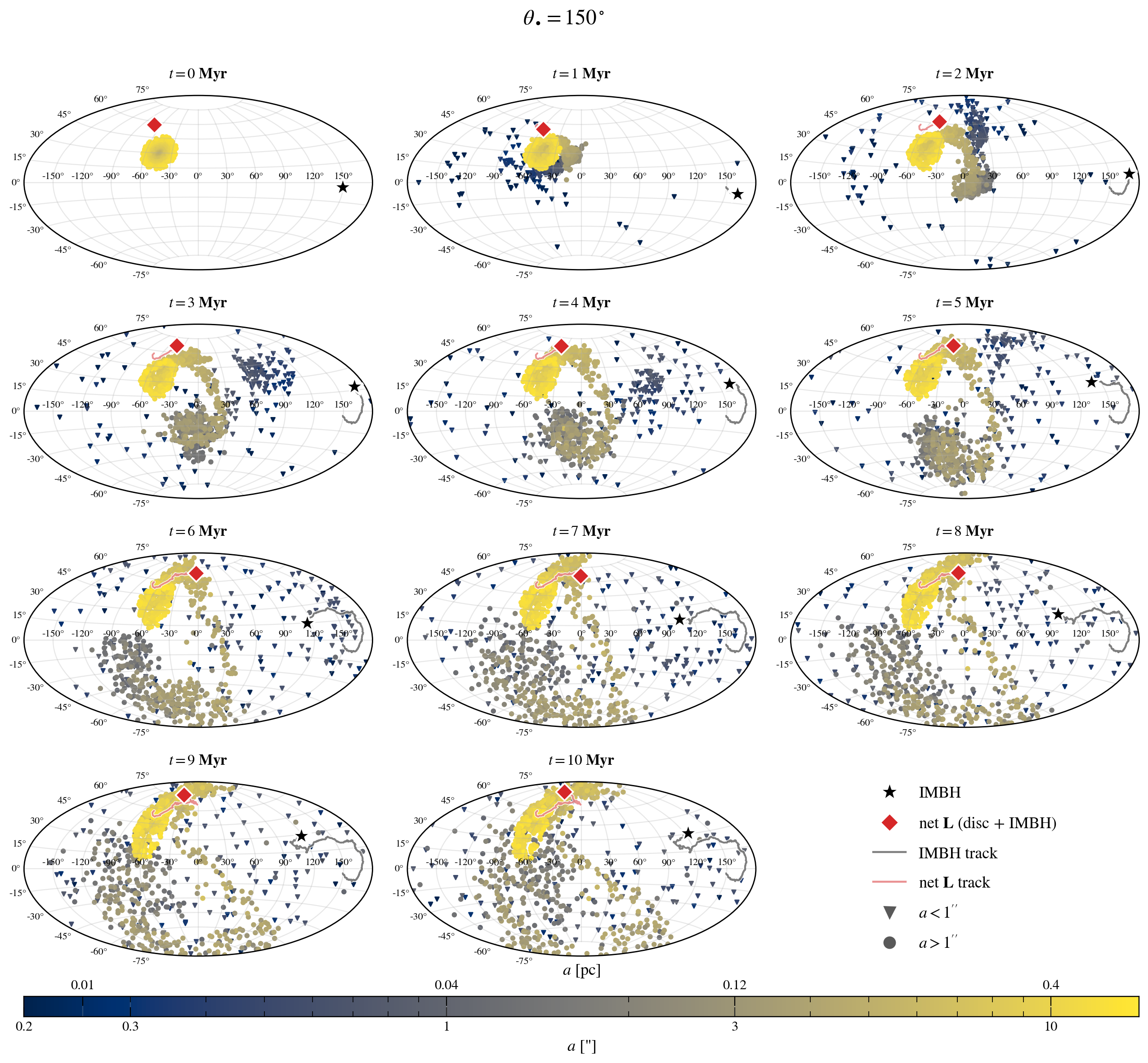}
    \caption{As Fig.~\ref{fig:skyev_noimbh}, for the fiducial retrograde
    $\theta_\IMBH=150^\circ$, $a_\IMBH=0.15$~pc, $m_\IMBH=2000\msun$ model.}
    \label{fig:skyev_i150}
\end{figure*}

Figures~\ref{fig:skyev_noimbh}--\ref{fig:skyev_i150} follow the Aitoff maps of
Fig.~\ref{fig:skymaps} through time, one model per figure, at $1$~Myr intervals
from $t=0$ to $10$~Myr: the model without an IMBH
(Fig.~\ref{fig:skyev_noimbh}), the $500\msun$, $a_\IMBH=0.05$~pc models
(Figs~\ref{fig:skyev_i45m500} and~\ref{fig:skyev_i135m500}), and the
$2000\msun$, $a_\IMBH=0.15$~pc models of Table~\ref{tab:runs}
(Figs~\ref{fig:skyev_i45}, \ref{fig:skyev_i135}, \ref{fig:skyev_i90}
and~\ref{fig:skyev_i150}). The projection, rotation
and colour scale are those of Fig.~\ref{fig:skymaps}, so the panels may be
compared directly with it; the black star marks the IMBH and downward
triangles the innermost stars ($a\le0.04$~pc, $1''$). The red diamond gives
the net angular-momentum direction of the disc and the IMBH together, and the
two faint tracks trace the paths that this net direction and the IMBH itself
follow from $t=0$ up to the epoch of each panel.

The sequences show the disc leaving its initial single, compact clump
(identical in every model) and the innermost stars dispersing across the sky
within a few Myr in every model, including the one without an IMBH. The
$2000\msun$ retrograde and polar models break the disc body into distinct
concentrations over the radial range the IMBH sweeps, and these separations
keep growing as the IMBH orbit decays. The prograde models instead retain a
single, progressively broadened concentration, with which the IMBH aligns, as
expected from the inclination evolution of Section~\ref{subsec:imbh-inc}; in
the low-mass $500\msun$ models, whose orbits barely decay, the disc likewise
broadens without separating into lasting components. Throughout, the outermost
part of the disc, at the yellow end of the colour scale, stays coherent.

\clearpage

\clearpage
\onecolumn
\section{Relaxation of the disc stars}
\label{app:disc-relaxation}

\begin{figure*}
\centering
\includegraphics[width=0.8\linewidth]{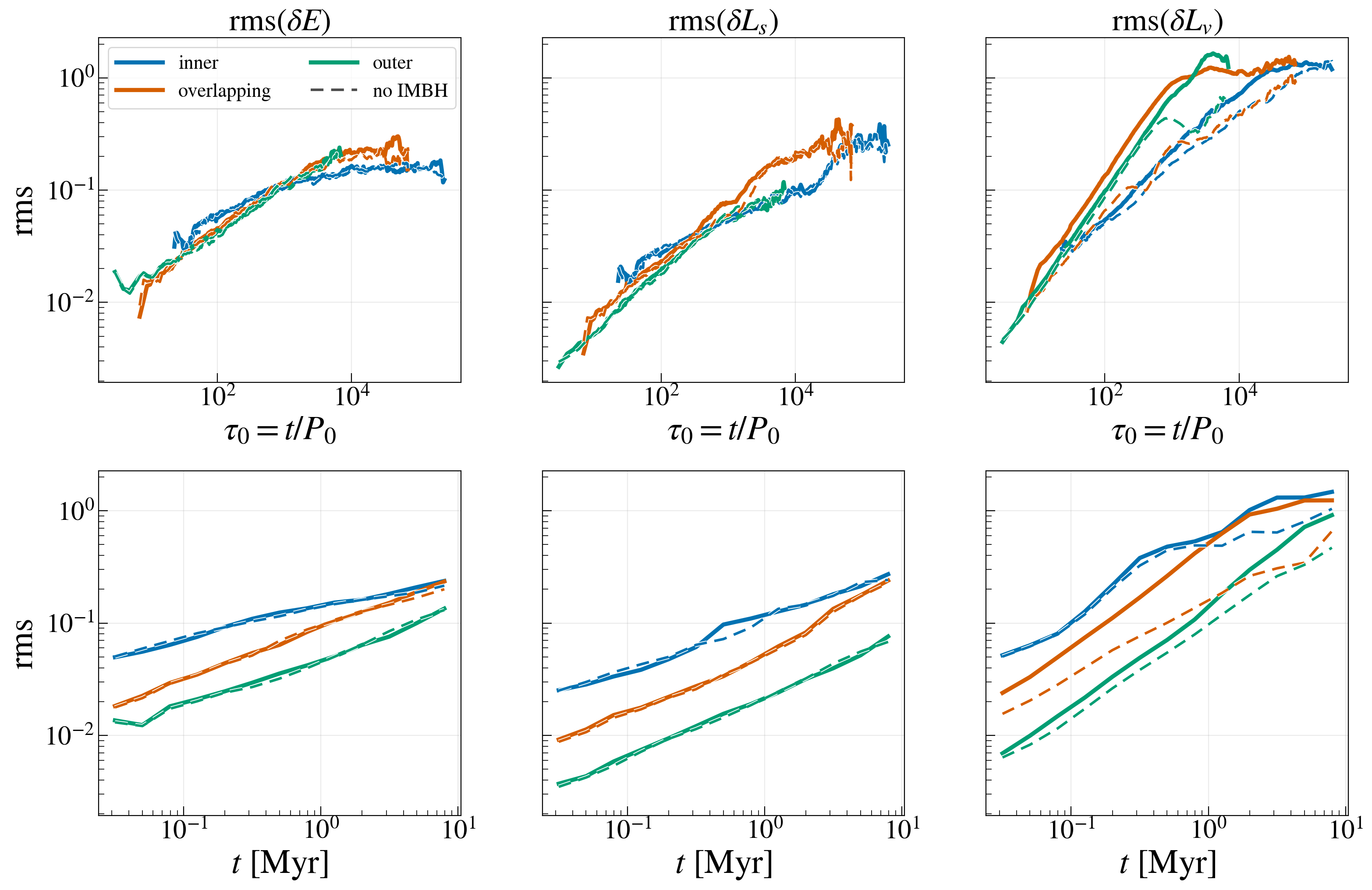}
\caption{Root-mean-square changes since $t=0$ in energy (left), angular-momentum magnitude (middle) and angular-momentum vector (right), normalised as in Eq.~\eqref{eq:disc-relaxation}. Thick solid curves show the fiducial $\theta_\IMBH=150^\circ$ model and dashed curves the model without an IMBH. Colours distinguish the inner (blue), overlapping (orange) and outer (green) groups. The upper row uses time in initial orbital periods, $\tau_0=t/P_0$; the lower row uses logarithmic bins of physical time, with equal weight per star.}
\label{fig:disc-relaxation}
\end{figure*}

\begin{multicols}{2}
We measure the changes in the disc stars' orbits following \citet{Panamarev2022}. For each star, we compute
\begin{equation}
\begin{aligned}
\delta E &= \frac{|E-E_0|}{|E_0|}, \qquad
\delta L_s = \frac{|J-J_0|}{J_{\mathrm{c},0}}, \\
\delta L_v &= \frac{|\boldsymbol{L}-\boldsymbol{L}_0|}{J_0},
\end{aligned}
\label{eq:disc-relaxation}
\end{equation}
where $E$ and $\boldsymbol{L}$ are the specific Keplerian energy and angular momentum, $J=|\boldsymbol{L}|$, and subscript $0$ denotes the initial orbit. The quantity $J_{\mathrm{c},0}$ is the specific angular momentum of a circular orbit with the same initial semi-major axis. We take the root mean square over stars in bins of $\tau_0=t/P_0$, where $P_0$ is each star's initial orbital period. We also bin the changes in physical time, averaging the squared changes of each star within a bin before averaging over stars, so each star has equal weight.

We use the orbital groups defined by Eq.~\eqref{eq:orb-cond}, with the fiducial IMBH orbit setting the boundaries in both runs. We retain measurements of the same stars in normalised-time bins shared by both runs. Each plotted bin contains at least ten stars.

Figure~\ref{fig:disc-relaxation} shows that the IMBH primarily increases the changes in the angular-momentum vector. In the time bin centred near $5$~Myr, these changes are larger by factors of $1.6$, $3.6$ and $2.2$ for the inner, overlapping and outer groups, respectively. The enhancement persists in the final bin, covering $6.0$--$10$~Myr. The energy and angular-momentum-magnitude changes differ by less than $20$ per cent between the runs in these late-time bins. 
\end{multicols}

\end{document}